\documentclass[11pt]{article}

\usepackage{amssymb,amsmath,latexsym,bm,amsfonts}
\usepackage{slashed}
\usepackage{graphicx}
\usepackage{longtable}
\usepackage{color,xcolor}
\usepackage{indentfirst}
\usepackage{subfigure}
\usepackage[margin=1in]{geometry}
\usepackage{mathtools,tensor}
\usepackage{footmisc} 

\usepackage{breqn}
\usepackage{authblk}
\usepackage{etoolbox}

\numberwithin{equation}{section}
\usepackage{stmaryrd}

\usepackage[nottoc ]{tocbibind}

\newcommand{\be}{\begin{equation}}
\newcommand{\ee}{\end{equation}}
\newcommand{\beqn}{\begin{eqnarray}}
\newcommand{\eeqn}{\end{eqnarray}}
\newcommand{\cD}{\mathcal D}

\newcommand{\cA}{\mathcal{A}}
\newcommand{\cE}{\mathcal{E}}
\newcommand{\cM}{\mathcal{M}}

\newcommand{\p}{\partial}

\DeclareMathOperator{\Real}{Re}

\DeclareMathOperator{\Imag}{Im}

\usepackage{breqn}
\usepackage[colorlinks=true,filecolor=black,citecolor=blue ]{hyperref}
\usepackage{cleveref}

\usepackage{eufrak}
 \usepackage{mathrsfs}
 \usepackage{mathrsfs}

\usepackage{mathtools}
\usepackage{latexsym,graphicx,verbatim,amssymb,amsfonts,amscd,amsbsy}

\def\bea{\begin{array}}
\def\bem{\begin{displaymath}}
\def\beq{\begin{equation}}

\def\eea{\end{array}}
\def\eem{\end{displaymath}}
\def\eeq{\end{equation}}

\def\Im{\mathop{\rm Im}}

\def\ov{\overline}

\def\Re{\mathop{\rm Re}}

\def\s2w{\sin^2 \theta_W}

\relax
\def\a{{\alpha}}
\def\b{{\beta}}
\def\dalpha{{\dot\alpha}}
\def\ad{{\dot\alpha}}

\def\bd{{\dot\beta}}

\def\crbig{\\\noalign{\vspace {3mm}}}
\def\bigint{{\displaystyle\int}}

\def\Fint{\bigint d^2\theta\,}
\def\Fbarint{\bigint d^2\ov\theta\,}
\def\Dint{\bigint d^2\theta d^2\ov\theta\,}

\renewcommand{\bea}{\begin{eqnarray}}
\renewcommand{\eea}{\end{eqnarray}}

\newcommand{\pa}{{\partial}}

\newcommand{\cN}{{\cal N}}
\newcommand{\q}{{\theta}}
\newcommand{\qb}{{\overline{\theta}}}
\newcommand{\hf}{\frac{1}{2}}

\def\a{\alpha}
\def\b{\beta}

\def\d{\delta}

\def\g{\gamma}

\def\l{\lambda}

\def\q{\theta}

\def\s{\sigma}

\newcommand{\de}{{\nabla}}
\newcommand{\deb}{{\overline{\nabla}}}

\def\ri{{\rm i}}

\newcommand{\bsubeq}{\begin{subequations}}
\newcommand{\esubeq}{\end{subequations}}

\newcommand{\bL}{{\mathbf L}}

\newcommand{\bB}{{\mathbf B}}

\newcommand{\bV}{{\mathbf V}}
\newcommand{\bW}{{\mathbf W}}
\newcommand{\bUp}{{\mathbf{\Upsilon}}}

\newcommand{\cDB}{\ov{\mathcal D}}

\relax
\def\baselinestretch{1}
\newcommand{\cL}{\mathcal  L} 
\newcommand{\cF}{\mathcal  F}

\newcommand{\cSp}{\Upsilon}   
\newcommand{\cH}{E}    
\newcommand{\cB}{\mathcal  B}    
 
\begin{document}

\renewcommand{\baselinestretch}{1.2}
\setlength{\parindent}{2em}
\setlength{\parskip}{0.4em}

\begin{titlepage}
  \ \\
 
 \hspace{12cm}\today
  
  \vskip 1cm

  \begin{center}
    
 { \Huge  \textbf{\textsf{Magnetic Deformation of \\[8pt]
 Super-Maxwell Theory in Supergravity
}}
 }

    \vspace{0.6 cm}
     
    {\Large  Ignatios Antoniadis~$^{a,b}$, Jean-Pierre Derendinger~$^{a}$,   Hongliang   Jiang~$^{a}$ \\ \medskip  and Gabriele
Tartaglino-Mazzucchelli~$^{a,c}$}

    \vspace{0.5cm}
    
    \normalsize{\it   
 $^{a}$Albert Einstein Center, Institute for Theoretical Physics, University of Bern, \\
 			Sidlerstrasse 5, 3012 Bern, Switzerland   \\[10pt]
			
 $^{b }$Laboratoire de Physique Th\'eorique et Hautes Energies - LPTHE,
Sorbonne Universit\'e, \\  CNRS, 4 Place Jussieu, 75005 Paris, France \\ [10pt]

	$^{c}$School of Mathematics and Physics, University of Queensland St Lucia, \\
	Brisbane, Queensland 4072, Australia
	}

    \vspace{0.5 cm}

{\texttt{antoniadis@itp.unibe.ch}}, \quad{\texttt{derendinger@itp.unibe.ch}},
 \quad {\texttt{jiang@itp.unibe.ch}},   {\texttt{g.tartaglino-mazzucchelli@uq.edu.au}}

    \vspace{0.5 cm}

\begin{abstract} \noindent
A necessary condition for partial breaking of ${\cal N}=2$ global  supersymmetry is the presence of nonlinear deformations 
of the field transformations which cannot be generated by background values of auxiliary fields. This work studies the simplest of these deformations which already occurs in ${\cal N}=1$ global supersymmetry,
and its coupling to supergravity. It can be viewed as an imaginary constant shift of the $D$-auxiliary real 
field of an abelian gauge multiplet. We show how this deformation describes the magnetic dual of a Fayet-Iliopoulos term,
a result that remains valid in supergravity, using its new-minimal formulation.
Local supersymmetry and the deformation induce a positive cosmological constant. 
Moreover, the deformed $U(1)$ Maxwell theory coupled to supergravity describes 
upon elimination of the auxiliary fields the gauging of $R$-symmetry, realised 
by the Freedman model of 1976. 
To this end, we construct the chiral spinor multiplet in superconformal tensor calculus by working out explicitly its transformation rules and use it for an alternative description of the new-minimal supergravity coupled to a $U(1)$ multiplet.
We also discuss the deformed Maxwell theory in curved superspace.

\end{abstract}
\vspace{1cm}
\end{center}
  
\end{titlepage}

\pagestyle{plain}
\pagenumbering{arabic}

 {\hypersetup{linkcolor=black}
 \tableofcontents
}

\newpage

\section{Introduction}

\noindent
Deformations of supersymmetry transformations play an important role for realising a partial breaking of extended supersymmetry~\cite{APT, ADtM, ADM}.
In ${\cal N}=2$ super-Maxwell theory, such deformations involve six parameters. 
Three of them can be generated by background values of auxiliary fields of the off-shell 
representation, in a real $SU(2)$ $R$-symmetry vector. The other three are their magnetic counterparts, absent in off-shell
representations. They can be obtained formally by considering a constant imaginary part for every component of the $SU(2)$ triplet of auxiliary fields. In other words, these six deformation parameters
form a complex $SU(2)$ vector $\vec Y$ and global supersymmetry is partially broken
${\cal N}=2\to{\cal N}=1$ if the deformation vector is non-trivial and nilpotent in the vacuum:
$$
| \vec Y|^2>0 \qquad;\qquad \vec Y^2 =0\,.
$$
This is explained for instance in Refs.~\cite{ADM,Antoniadis:2019gbd}. 

An interesting question is to understand the coupling of a deformed supersymmetric theory to supergravity in relation to (partial) supersymmetry breaking.
In this work, we make a first step towards this investigation by studying a non-trivial supersymmetry deformation 
in a simpler context, namely at the level of ${\cal N}=1$.
The type of deformation considered in the present paper has already appeared 
in the context of the supersymmetric Dirac-Born-Infeld theory and partial ${\cal N}=2\to{\cal N}=1$ 
supersymmetry breaking
\cite{K,Antoniadis:2019gbd,Antoniadis:2019xwa}
but here we will focus on a purely ${\cal N}=1$ analysis.

Indeed, in ${\cal N}=1$ super-Maxwell theory, 
the real auxiliary field $D$ can generate
an ``electric" deformation (equivalent to a Fayet-Iliopoulos term), while its
magnetic counterpart, which can be formally obtained by adding to $D$ a constant imaginary part in the supersymmetry variations,
corresponds to an integration constant in the supersymmetric Bianchi identity. In our analysis here, we show that such an integration constant is equivalent to a `magnetic' Fayet-Iliopoulos term, dual under electric-magnetic (EM) duality to an `electric' Fayet-Iliopoulos term. A corollary of this result is that one cannot add `electrically' charged chiral multiplets in a local action containing the deformation, since they would correspond to magnetic monopoles 
in the dual theory which has a Fayet-Iliopoulos term. In the presence of several $U(1)$'s with corresponding deformation parameters, charged matter should satisfy the condition of being neutral under the $U(1)$ combination containing the deformation, while non trivial charges can exist with respect to all orthogonal combinations for which supersymmetry variations are not deformed. 

We then proceed to the description of the deformation in supergravity, considering the simplest case of pure 
${\cal N}=1$ supergravity coupled to an abelian $U(1)$ multiplet. The main observation is that an integration constant in the supersymmetric Bianchi identity can be obtained as a background value of a linear multiplet. Since the Bianchi identity involves the chiral spinor gauge field-strength superfield $W_\alpha$, it is natural to consider a general chiral spinor superfield which contains the degrees of freedom of a Maxwell multiplet and of a linear multiplet~\cite{dWR, KU}. The usual Bianchi identity eliminates the latter and leaves the gauge multiplet in the physical spectrum, while the presence of an integration constant may arise from a background value for the linear multiplet. To avoid adding extra degrees of freedom in the theory, we identify the linear multiplet with the compensator of the new-minimal \cite{SW} off-shell formulation of ${\cal N}=1$ supergravity~\cite{dWR, KU, FGKVP}. 
The deformed Maxwell theory can thus be constructed in a similar way as the undeformed one by implementing the deformation in the Bianchi identity of  $W_\alpha$  stemmed from the linear compensator. 

As in the global case,  one finds that the deformation becomes a Fayet-Iliopoulos term in supergravity after performing a EM duality. The Fayet-Iliopoulos term generates  a positive cosmological constant proportional to the square of the $U(1)$ coupling and can be described by  the well-known Freedman model where the $R$-symmetry is gauged and  the gravitino and gaugino are charged under it~\cite{F}. Therefore, the deformed theory  provides a magnetic dual description of the Freedman model \emph{off-shell}.
However, after eliminating the auxiliary fields in the supergravity context on the deformed theory side, the leftover propagating $U(1)$ vector boson gauges again the $R$-symmetry under which the fermions are charged (see e.g.~\cite{FGKVP, FVP}). 

More precisely, 
from the argument in global supersymmetry presented above, one would expect the absence of charged particles on the side of the deformed theory. Indeed the fermions (gravitino and gaugino) are neutral under the off-shell Maxwell field. However, one finds that the latter becomes unphysical and upon appropriate gauge conditions and integration over all auxiliary fields, the physical propagating gauge field corresponds now to the $R$-symmetry $U(1)$ gauge field, and thus the fermions remain charged. The resulting theory is shown in fact to be
equivalent to the Freedman model \emph{on-shell}.

The outline of the paper is the following. In Section~2, we discuss the $D$-deformation in ${\cal N}=1$ global supersymmetry and show that it corresponds to a magnetic Fayet-Iliopoulos term. In Section~3, we shortly review
the new-minimal formulation of pure supergravity \cite{SW,dWR} needed in the following. In Section~4, we construct the coupling of the deformed super-Maxwell theory to supergravity using the method described above which consists of modifying the Bianchi identity of a chiral spinor superfield by a term proportional to the compensator linear multiplet with a constant coefficient playing the role of the integration constant in global supersymmetry. In Section~5, we describe the introduction of the Fayet-Iliopoulos term in new-minimal 
supergravity and show how this formulation is related to an old-minimal theory with the Maxwell multiplet used to gauge the $U(1)_R$ superconformal symmetry, following for instance ref.~\cite{FGKVP}. This is the Freedman model \cite{F}. We then compare it with the deformed super-Maxwell theory and verify that the two theories are indentical once auxiliary fields have been eliminated. In Section~6, we work out the EM duality in supergravity, generalising the result of global supersymmetry and we show that in the absence of matter the deformed super-Maxwell theory is dual to the Freedman model of gauged $R$-symmetry. Section~7 contains a summary of our results and an outlook. 
The paper is also accompanied by three Appendices:
Appendix~A contains our conventions and useful formulae; Appendix~B describes
the local superconformal chiral spinor multiplet used
in the main body of the paper;
Appendix~C is devoted  to a complementary 
description of the deformed Maxwell theory in curved superspace.

\section{$D$-deformation in global \boldmath{${\cal N}=1$} supersymmetry}
\label{global-susy}

\noindent
The ${\cal N}=1$ super-Maxwell theory is usually formulated in terms of a chiral spinor superfield 
$W_\alpha$ subject to the superfield condition\,\footnote{
We use covariant derivatives 
$$
D_\alpha=\frac{\p}{\p \theta^\alpha} - i (\sigma^\mu \ov\theta)_\alpha \p_\mu, \qquad
\ov  D_{\dot\alpha}=\frac{\p}{\p \ov\theta^{\dot \alpha} } -  i ( \theta \sigma^\mu   )_{\dot \alpha} \p_\mu. 
$$
}
\beq
\label{Bianchi}
D^\alpha W_\alpha = \ov D_\dalpha \ov W^\dalpha
\eeq
which imposes the Bianchi identity $\partial_{[\mu}F_{\nu\rho]}=0$ on the Maxwell field-strength $F_{\mu\nu}$.

Consider instead an arbitrary chiral spinor superfield $\cSp_\alpha$, $\ov D_\dalpha\cSp_\beta=0$ 
and its conjugate 
$\ov \cSp_\dalpha=(\cSp_\alpha)^*$.
The superfield $D^\alpha \cSp_\alpha - \ov D_\dalpha \ov \cSp^\dalpha$ is real and linear. The condition
\beq
\label{defcSpL}
 D^\alpha \cSp_\alpha - \ov D_\dalpha \ov \cSp^\dalpha = L
\eeq
defines $\cSp^\alpha$ for a given $L$, $\ov{DD}L=0$, up to a solution $\cSp_{0\alpha}$ of eq.~(\ref{Bianchi}),
which is\,\footnote{With this convention,
$\ov W_\dalpha = - (W_\alpha{})^*$.}
\beq
\label{glob2}
\cSp_{0\alpha} = iW_\alpha, \qquad \ov \cSp_{0\dalpha} = i\ov W_\dalpha, \qquad
W_\alpha = -{1\over4}\,\ov{DD} D_\alpha V, \qquad \ov W_\dalpha = -{1\over4}\,DD\ov  D_\dalpha V,
\eeq
where $V$ is a real superfield. 
In the simplest case, we can take $L$ to be a (real) constant,
$L=4\zeta$. This amounts to give a supersymmetric-invariant background value to the lowest scalar 
component of $L$. The solution of 
\beq
\label{defcSp}
D^\alpha \cSp_\alpha - \ov D_\dalpha \ov \cSp^\dalpha = 4\, \zeta
\eeq
is
\beq
\label{glob4}
\cSp_\alpha = -\zeta\, \theta_\alpha + iW_\alpha \equiv i W^{\rm def}_\alpha,
\qquad\qquad
\ov \cSp_\dalpha = - \zeta\, \ov\theta_\dalpha + i\ov W_\dalpha \equiv i\ov{W}^{\rm def}_\dalpha
\eeq
where the deformed chiral Maxwell superfield is
\beq
\label{glob5}
W^{\rm def}_\alpha  = -i\lambda_\alpha + \theta_\alpha (D + i \zeta)
- {i\over2}(\sigma^\mu\ov\sigma^\nu\theta)_\alpha F_{\mu\nu}
- \theta\theta\, (\sigma^\mu\partial_\mu\ov\lambda)_\alpha
\eeq
and satisfies the \emph{deformed} supersymmetric Bianchi identity $D^\alpha W^{\rm def}_\alpha  - \ov D_\dalpha
\ov{W}^{{\rm def}\,\dalpha} = -4i\, \zeta$. The supersymmetry variation of the gaugino is now
\beq
\label{glob6}
\delta\lambda_\alpha = -\zeta\epsilon_\alpha + i\,D\epsilon_\alpha
- {1\over4}\, \epsilon[\sigma^\mu,\ov\sigma^\nu] F_{\mu\nu}.
\eeq
Therefore $\lambda$ appears to transform like a Goldstino because of the deformation. 
\footnote{Note that the theory we consider here is  essentially free where the fermion enjoys the shift symmetry. Therefore the non-linear gaugino transformation in \eqref{glob6}  is of little physical significance. This section serves as a simple illustration of the deformation whose real  significance will be shown in supergravity. }
 
To see the role of the deformation, consider the lagrangian
\be\label{cspaction}
\cL=-\frac{1}{2} \Imag \Big[ \widetilde\tau \int d^2 \theta\, \cSp^2 \Big]- {1\over 2}\Dint U(D^\alpha \cSp_\alpha - \ov D_\dalpha \ov \cSp^\dalpha - 4\, \zeta),\quad\,
\widetilde\tau=\frac{i }{{\tilde g}^2}+\vartheta\,, 
\ee
where  $\cSp_\alpha$ is a chiral spinor superfield without extra constraint and  $U$ is a real scalar
superfield. 

Eliminating $U$ imposes the constraint \eqref{defcSp} on $\cSp_\alpha$ and leads to a deformed Maxwell theory
\beq
\label{MagMaxwell}
\cL_M= \frac{1}{2} \Imag \Big[ \widetilde\tau \int d^2 \theta\, W_{\rm def}^2 \Big]
= \frac{1}{2} \Imag \Big[ \widetilde\tau \int d^2 \theta\, W^2\Big]
+\frac{1}{2}  \Big(-\zeta^2 \Imag \widetilde\tau +2 \zeta D \Real \widetilde\tau \Big), 
\eeq
where we used $W^{\rm def}_\alpha =W_\alpha + i\zeta\,\theta_\alpha$.
Notice the emergence of a Fayet-Iliopoulos term proportional to the theta-angle $\Real\widetilde\tau$, induced by the deformation, as was noticed in the context of ${\cal N}=2$~\cite{ADM, Antoniadis:2019gbd}.

Alternatively, we can integrate by parts and rewrite the lagrangian~\eqref{cspaction} as
\be
 \cL=-\frac{1}{2} \Imag   \int d^2 \theta\, \Big(\widetilde\tau\,\cSp^2 +  \frac{ i}{2} 
 \cSp^\alpha  \ov{DD}D_\alpha U  \Big)
 + 2 \,\zeta \Dint U .
\ee
Eliminating $\cSp$, we arrive at 
\be\label{EleMaxwell}
 \cL_E= \frac{1}{2} \Imag \left[ \tau  \int d^2 \theta\, W^{\,2}\right]+ 2\zeta \Dint U\quad, \quad
 W_\alpha = -{1\over4}\,\ov{DD}D_\alpha U\quad;\quad \tau=-{1\over\widetilde\tau}\,.
\ee

The last lagrangian \eqref{EleMaxwell} is the standard ``electric" expression of the super-Maxwell theory with a Fayet-Iliopoulos term.
It is the electric dual of the ``magnetic" lagrangian \eqref{MagMaxwell}, with the deformation induced by $\zeta$.
In both  the electric and magnetic descriptions, a constant value 
\beq
\label{constant}
- \frac{\zeta^2}{2} {\widetilde\tau\ov{\widetilde\tau}\over\Im\widetilde\tau} = - \frac{\zeta^2}{2} {1\over\Im \tau}
\eeq
is  added to the super-Maxwell lagrangian, after eliminating $D$. 
The Fayet-Iliopoulos term can be seen as a nonlinear deformation of $\delta\lambda$ induced by
a constant $\langle D\rangle$. In the magnetic dual, the deformation is induced by the parameter 
$\zeta$ introduced in the Bianchi identity  \eqref{defcSp}. As expected, the electric and magnetic couplings
are related by $\tau \leftrightarrow -1/\tau\equiv \widetilde\tau$. At the superfield level,
the magnetic \eqref{MagMaxwell} and the electric \eqref{EleMaxwell} versions 
are related by the map of field strengths
$W_\alpha\longleftrightarrow \widetilde\tau \, W^{\rm def}_\alpha$.\,\footnote{
See ref.~\cite{K, KT} for more detail about electric-magnetic duality transformations in 
superspace with and without Fayet-Iliopoulos term and Bianchi identity deformations.}

To summarise, the above argument shows that the deformation in the Bianchi identity is equivalent to 
a Fayet-Iliopoulos term in the `magnetic' dual theory. The 
constant term \eqref{constant}, which is
irrelevant in the context of global supersymmetry, 
acquires relevance when the theory is coupled to ${\cal N}=1$ supergravity or if $\tau$ is promoted
to an analytic function of some neutral chiral superfields. It is 
then interesting to evaluate the scalar potential
obtained in the presence of both `electric' and `magnetic' Fayet-Iliopoulos terms. This is possible by adding in the deformed theory \eqref{MagMaxwell} a term linear in the auxiliary field $D$ with a constant coefficient $\xi$, which transforms with a total derivative under supersymmetry. In this case, the constant \eqref{constant} is replaced by 
the scalar potential
\beq\label{ccglobal}
V = {1\over 2}\left\{{(\xi+\zeta\Re\widetilde\tau)^2\over\Im\widetilde\tau}+\zeta^2\Im\widetilde\tau\right\}
\eeq
which is invariant under $\tau \leftrightarrow -1/\tau$ (or $\widetilde\tau \leftrightarrow -1/\widetilde\tau$) 
and the exchange $(\xi,\zeta)\to (\zeta,-\xi)$. This contribution can be obtained from an ${\cal N}=2$ supersymmetric theory by restricting the electric and magnetic triplets of Fayet-Iliopoulos terms to the $D$-direction and identify $\tau$ with the second derivative of the prepotential~\cite{ADM, Antoniadis:2019gbd}. Note that 
it has 
a runaway behaviour towards strong coupling $(\Im\tilde\tau=0)$ after minimisation with respect to the theta-angle $\vartheta=\Re\tilde\tau$, when minimising with respect to $\tilde g$. 
A non-trivial superpotential is needed to stabilise the theory.

Another question concerns the addition of charged matter. This is straightforward in the dual version of the theory containing a Fayet-Iliopoulos term. On the contrary, it is not possible to add charged matter in the presence of the deformation. One can infer this from the fact that the real vector superfield needed to describe gauge-invariant kinetic matter lagrangians cannot include the deformation.\footnote{In other words, 
$W^{\rm def}_\alpha$ in eq.~\eqref{glob4} cannot be written as $-{1\over4}\ov{DD}D_\alpha V$, with a real $V$.}
Consider the standard real and gauge invariant $\ov\Phi e^V \Phi$ used to couple a chiral multiplet 
(with $U(1)$ charge one) to the real Maxwell superfield $V$. The kinetic lagrangian includes a 
Yukawa interaction involving the gaugino:
\beq
\ov\Phi e^V \Phi \qquad\longrightarrow\qquad 
{i\over\sqrt2}\, \Bigl(\ov z \lambda\psi - z\ov{\lambda\psi}\Bigr) .
\eeq
Its deformed variation is 
\beq
{i\over\sqrt2}\, \Bigl(\ov z (\delta_{\rm def}\lambda)\psi - z{(\delta_{\rm def}\ov\lambda)\ov\psi} \Bigr)
= -{i\over\sqrt2}\zeta\, \Bigl(\ov z \epsilon\psi - z \ov{\epsilon\psi} \Bigr)
=  -i\zeta( \ov z\delta z -z\delta\ov z),
\eeq
where $\delta_{\rm def}\lambda_\alpha = -\zeta\,\epsilon_\alpha$, 
$\delta_{\rm def}\ov\lambda_\dalpha = -\zeta\,\ov\epsilon_\dalpha$. 
This variation cannot be cancelled by a $\zeta$--dependent counterterm added to the lagrangian. This
simple argument easily generalizes to an arbitrary K\"ahler potential.  

It follows that in the case of several $U(1)$'s with different deformation parameters $\zeta_i$, charged matter fields should satisfy the requirement of being neutral under the $U(1)$ combination containing the deformation, while non trivial charges can exist only with respect to the remaining orthogonal combinations for which supersymmetry variations are not deformed. This requirement is translated to the follwing condition on the physical charges $q_i$:
\beq
\sum_i q_i\zeta_i=0\,,
\eeq
where the various gauge couplings are implicit in the definition of the charges.

This apparent obstruction seems to have an important physical implication for magnetic monopoles.
Indeed, states with magnetic charges may be in principle added in the theory with a deformation in the electric Bianchi identity
(although its local description is not known),
since they correspond to electrically charged states in the dual theory with a Fayet-Iliopoulos term. 
On the other hand, magnetically charged states seem to be forbidden for a $U(1)$ with an electric Fayet-Iliopoulos term, since they would correspond to electrically charged matter in the deformed theory in view of the obstruction described above.

In the context of global supersymmetry without coupling to matter, which is the focus of our paper,
the above discussion is not fruitful: we are merely
considering a free theory. It acquires relevance when coupled to ${\cal N}=1$ supergravity that we study in the following.
 
\section{Supergravity}

\noindent
In the previous section, we have introduced in global supersymmetry a deformation of super-Maxwell 
theory which is the magnetic dual of the standard, electric, Fayet-Iliopoulos term. 
In the rest of the paper, we extend this construction to supergravity. 
We use a superconformal formulation\footnote{Reviewed in ref.~\cite{FVP}.}, 
which is certainly appropriate to
describe the super-Maxwell system, and, since the idea is to use the linear multiplet $L$ of 
eq.~\eqref{defcSpL} as compensating multiplet, we use the new-minimal formulation of ${\cal N}=1$
supergravity \cite{SW}.\,\footnote{By definition, minimal supergravity has an off-shell multiplet of fields with the gravitino 
as the sole fermion. It has then 12 bosonic and 12 fermionic component fields ($12_B+12_F$). There are two 
choices of auxiliary fields, leading to old- \cite{old-minimal} and new-minimal \cite{SW} supergravity.}
As usual, the resulting new-minimal theory can be transformed (before Poincar\'e 
gauge fixing) into the old-minimal one by a superconformal chiral-linear duality transformation
\cite{FGKVP}.

The superconformal formulation of new-minimal ${\cal N}=1$ supergravity uses a real linear multiplet $L$ as 
compensator \cite{dWR, KU, FGKVP}. 
Its lowest component $C$ is used to gauge-fix Weyl symmetry with the condition
$C = \kappa^{-2}$. Hence, one can transport the discussion of global supersymmetry to new-minimal
supergravity by defining a chiral spinor multiplet with weights $w=n=3/2$\,\footnote{
We denote by $w$ and $n$ the Weyl
weight and the $U(1)_R$ charge of a field or a multiplet. The normalization of $U(1)_R$ is such that
$w=n$ for (the lowest component of) a chiral multiplet.
Note that, we mostly follow the conventions of ref.~\cite{FVP}, with some exceptions stated in 
Appendix~\ref{AppA}.}
through the equation\,\footnote{
By $D^\alpha \cSp_\alpha$ or 
$\ov D_\dalpha \ov \cSp^\dalpha$, we mean the local superconformal multiplets with weights $w=2$, $n=0$ 
corresponding to the global superfields $D^\alpha \cSp_\alpha$ or $\ov D_\dalpha \ov \cSp^\dalpha$.
A curved superspace meaning to the covariant derivatives 
$D^\alpha$ and $\ov D_\dalpha$ can be given 
by using a ``conformal superspace'' approach 
\cite{Butter} to ${\cal N}=1$ conformal supergravity,
see Appendix \ref{superspace}.}
\beq
\label{localcSpL}
D^\alpha \cSp_\alpha - \ov D_\dalpha \ov \cSp^\dalpha = 4\zeta \, L.
\eeq
As a consequence, $\cSp_\alpha$ includes the Maxwell multiplet and the ``prepotential" of the compensating 
real linear multiplet $L$. Gauge-fixing superconformal symmetry
will then generate a deformation parameter $4\zeta\kappa^{-2}$ in the 
Poincar\'e theory, completely analogous to $4\zeta$ in the global case.

\subsection{New minimal supergravity}
\label{new-minimal_sugra-0}

\noindent
The auxiliary fields of the $12_B+12_F$ off-shell multiplet of new-minimal supergravity are
\cite{SW} an antisymmetric tensor 
$ B_{\mu\nu}$ and a vector field $A_\mu$. Both are gauge fields and contain   $3_B+3_B$ field components.
The vector $A_\mu$ is the superconformal $U(1)_R$ gauge field, and $ B_{\mu\nu}$ is in the linear compensating multiplet $L$ together with the real scalar $C$ and Majorana fermion $\chi$.  It is 
convenient to describe the linear multiplet $L$ as a real multiplet with weight $w=2, n=0$ and components:
\be\label{newL}
L=\Big( C,\quad \chi,\quad   0, \quad -\cH_a, \quad -\gamma^b D_b \chi,  \quad -\Box C\Big) , \qquad\qquad
D^a \cH_a=0.
\ee
The constraint $D^a \cH_a=0$ can be solved explicitly in terms of $B_{ab}$ and the solution is given in eq.~\eqref{His1}. 
The superconformal construction \cite{dWR} of the new-minimal theory \cite{SW} is based on
the lagrangian
\beq
\label{newSG}
{\cal L}_{\text{new-min}}  = {3\over2}\left[L \ln {L\over\ov S   S} - L\right]_D ,
\eeq
where $S$ is a non-dynamical chiral multiplet with $w=n=1$ and $[\ldots]_D$ is the real invariant density formula
for a real multiplet with $w=2$.\footnote{The term linear in $L$ is added for convenience: it only contributes
with a derivative.} Since $[L(\Lambda+\ov\Lambda)]_D$ is a derivative for a chiral multiplet $\Lambda$ with
$w=n=0$, the action is invariant under the gauge transformation
\beq
S \qquad\longrightarrow\qquad e^\Lambda S
\label{gaugeS}
\eeq
even if this gauge symmetry is not explicitly gauged.\footnote{
An explicit gauging for the chiral multiplet is given by   $\ov S e^V S$ with $V$ being the gauge multiplet.}
Since the zero-weight real multiplet
$\ln {L\over\ov SS}$ transforms  under (\ref{gaugeS}) as a Maxwell vector multiplet, one can take the Wess-Zumino gauge
by an appropriate choice of the components $\varphi$, $\psi_L$ and $f$ of the chiral multiplet $S$. In terms 
of the components $C$, $\chi$ and $ B_{\mu\nu}$ of $L$,
the choice is
\beq 
\varphi = \sqrt C e^{i\alpha}, \qquad \psi_L = {i\over\sqrt{2C}} e^{i\alpha} \chi_L, \qquad f=0
\eeq
which as usual leaves arbitrary the phase $\alpha$ of $\varphi$ which always appears in the combination 
$A_\mu-\partial_\mu\alpha$. It thus can be eliminated with a $U(1)_R$ gauge transformation. 
Note that this procedure respects the superconformal $U(1)_R$ symmetry.
Poincar\'e gauge fixing applied to the components $C$, $\chi$ and $ B_{\mu\nu}$ of $L$ is then
\beq\label{newgaugefix}
C = {1\over\kappa^2} \qquad \makebox{(Weyl)}, \qquad\qquad \chi=0 \qquad \makebox{(${\cal S}$ supersymmetry)}.
\eeq
In addition, conformal boosts are fixed by the choice $b_\mu=0$ (Weyl gauge field).
With these choices and gauge fixings for $S$ and $L$, one can easily use the $D$-density 
formula~\eqref{DtermFormla} to compute the component expansion of the lagrangian \eqref{newSG} and
obtain the pure new-minimal Poincar\'e supergravity off-shell theory.
These gauge-fixing conditions will be used in the rest of this section.

There are two useful expressions of the new-minimal Poincar\'e theory. The first is  
\beq\label{minSG1}
\begin{array}{rcl}
{\cal L}_{\text{new-min}} &=& \displaystyle
{e\over2\kappa^2}\,[ {  R} - \ov\psi_\mu\gamma^{\mu\nu\rho}{\cal D}_\nu\psi_\rho ] 
- {3\kappa^2\over 4}\,e \cH^a\cH_a - {3\over2}\,  e \, A_\mu \epsilon^{\mu\nu\rho\sigma}\partial_\nu  B_{\rho\sigma} ,
\end{array}
\eeq
where $R = e^\mu_a  e^\nu_b R_{\mu\nu}^{ab}(\omega(e,\psi))$,
\beq\label{EaBab}
\cH^\mu =  {1\over2}\epsilon^{\mu\nu\rho\sigma}\partial_\nu B_{\rho\sigma}
+ {1\over4\kappa^2}\, \epsilon^{\mu\nu\rho\sigma}\ov\psi_\rho\gamma_\sigma\psi_\nu
\eeq
and
\beq
{\cal D}_\nu\psi_\rho = \cD_\nu^{(P)}\psi_\rho - {3\over2} i A_\nu\gamma_5\psi_\rho, 
\qquad\qquad
\cD^{(P)}_\nu\psi_\rho = \partial_\nu\psi_\rho + {1\over8} \, \omega_{\nu ab}(e,\psi) 
[\gamma^a,\gamma^b]\psi_\rho .
\eeq
Note for future use that eq.~\eqref{EaBab} solves the constraint
\beq
\label{Eaconstraint}
\partial_\mu \Bigl(eE^\mu 
- {ie\over4\kappa^2}\, \ov\psi_\nu\gamma^{\mu\nu\rho}\gamma_5\psi_\rho\Bigr)=0 ,
\eeq
where the second term includes the divergence of the gravitino chiral current.
This first form \eqref{minSG1} of the new-minimal action is explicitly invariant under $U(1)_R$,
with lagrangian variation 
\beq
\delta_R\, {\cal L}_{\text{new min}}  = - {3\over2}\,\partial_\mu\left[ 
e\,\lambda_T\,\epsilon^{\mu\nu\rho\sigma}\partial_\nu  B_{\rho\sigma}\right]
\eeq
induced by the $A_\mu$ term is eq.~\eqref{minSG1}.

The second equivalent expression useful to eliminate auxiliary fields is 
\beq\label{newSG2}
\begin{array}{rcl}
{\cal L}_{\text{new-min}} &=& \displaystyle
{e\over2\kappa^2}\,[ {  R} - \ov\psi_\mu\gamma^{\mu\nu\rho}{\cal D}^{(P)}_\nu\psi_\rho ] 
- {3\kappa^2\over 4}\,e \cH^a\cH_a  - 3e\, A_a\cH^a
\crbig
&=& \displaystyle
{e\over2\kappa^2}\,[ {  R} - \ov\psi_\mu\gamma^{\mu\nu\rho}{\cal D}^{(P)}_\nu\psi_\rho ] 
 - 3e\,  A'_a\cH^a ,
\end{array}
\eeq
defining a shifted $U(1)_R$ gauge field as
\beq
\label{Aredef}
 A'_a = A_a + {\kappa^2\over4}\, \cH_a.
\eeq
In pure supergravity, the equations of motion of the auxiliary fields lead to $\cH_a=0$ and 
$A_a= A'_a=$ \emph{pure gauge}, which then also vanishes by a gauge choice.

The spin connection $\omega_\mu{}^{ab}=\omega_\mu{}^{ab}(e,\psi)$ 
solves the constraint $R_{\mu\nu}^a(P)=0$ on the curvature of space-time translations, with $b_\mu=0$. It decomposes into
\beq
\label{spinconn}
\omega_\mu{}^{ab} =\omega_\mu{}^{ab}(e,\psi)= \omega_\mu{}^{ab}(e) + \kappa_\mu{}^{ab},
\eeq
with Poincar\'e spin connection
\beq\label{omegae}
\omega_\mu{}^{ab}(e) =
- {1\over2}(\partial_\mu e_\nu^a - \partial_\nu e_\mu^a) e^{\nu b}
+ {1\over2}(\partial_\mu e_\nu^b - \partial_\nu e_\mu^b) e^{\nu a}
- {1\over2}e^{\rho a}e^{\sigma b}(\partial_\rho e_\sigma^c - \partial_\sigma e_\rho^c) e_{\mu c} 
\eeq
and contorsion tensor 
\beq
\label{contor}
\kappa_\mu{}^{ab} =
{1\over4}\Bigl[ \ov\psi_\mu\gamma^a\psi^b + \ov\psi^a\gamma_\mu\psi^b - \ov\psi_\mu\gamma^b\psi^a \Bigr].
\eeq
When the theory is expressed in terms of the Poincar\'e spin connection $\omega_\mu{}^{ab}(e)$, 
the kinetic supergravity lagrangian produces the usual four-gravitino interactions
\beq
\label{4grav}
{\cal L}_{\rm 4,SG} =
{e\over16\kappa^2}\, \Bigl[ 4(\ov\psi_\mu\gamma^\mu\psi_\rho)(\ov\psi_\nu\gamma^\nu\psi^\rho)
- (\ov\psi_\mu\gamma_\nu\psi_\rho)(\ov\psi^\mu\gamma^\nu\psi^\rho)
- 2 (\ov\psi_\mu\gamma_\nu\psi_\rho)(\ov\psi^\mu\gamma^\rho\psi^\nu) \Bigr].
\eeq

\section{The deformed super-Maxwell theory in supergravity}  \label{deformedL}

\noindent
The description of the deformed Maxwell theory in conformal supergravity uses a full chiral spinor multiplet 
$\cSp$ with weight $w=n=3/2$ and with
$8_B+8_F$ off-shell field components. As seen in eq.~\eqref{localcSpL} which has a superconformal 
version and in Appendix \ref{AppB}, it combines the super-Maxwell fields and the linear multiplet $L$.
Then, as outlined in the global case, applying the Poincar\'e gauge-fixing conditions \eqref{newgaugefix} 
to $L$ used as compensating multiplet will generate a deformation in the Poincar\'e supergravity theory.

The field content of the chiral spinor superfield includes the real scalars $C$ and $D$, the two-form (non-gauge) 
field $\cB_{ab}$ and two Majorana spinors $\lambda$ (gaugino) and $\chi$.
Since this superconformal multiplet does not seem available in the literature, its supersymmetry 
variations are given in Appendix~\ref{AppB}. It includes two submultiplets: the Maxwell multiplet with fields 
$\lambda$, $D$, $\cA_a$ (with field-strength $\cF_{ab}$), and the linear multiplet with fields $C$, 
$\chi$ and the gauge field $B_{\mu\nu}$. Its decomposition into these two $4_B+4_F$ submultiplets is
consistent. In particular, the vector field $\cH^a$ present in the supersymmetry variations is defined as 
$\cH^a =  {1\over2}\, \epsilon^{abcd} \cD_b \cB_{cd}$ in the chiral spinor multiplet and as the solution of 
$\cD^a \cH_a=0$ in the linear multiplet, with the covariant derivatives relevant to each multiplet. That
both definitions lead to the same expression depending on $B_{\mu\nu}$, $C$ and $\chi$ follows from the 
Bianchi identity holding among components of the Maxwell submultiplet. This argument also implies the decomposition $\cB_{\mu\nu}=B_{\mu\nu}-\widehat \cF_{\mu\nu}$ with the superconformal Maxwell field-strength 
\be
\label{cFdef}
\widehat \cF_{\mu\nu} = \partial_\mu   \cA_\nu - \partial_\nu   \cA_\mu 
+ {1\over2}\, \ov\psi_\mu\gamma_\nu\lambda
- {1\over2}\, \ov\psi_\nu\gamma_\mu\lambda\,,
\ee
where we use the symbol hat to denote the superconformal field strength, including fermions.

To characterize the interaction of the linear submultiplet identified as the compensator of new-minimal supergravity with the super-Maxwell fields, we need a coupling constant $\zeta$.
It is simply introduced by rescaling the components of the linear multiplet
$$
C, \quad\chi, \quad B_{\mu\nu} \qquad\longrightarrow\qquad 
\zeta C, \quad \zeta \chi, \quad \zeta B_{\mu\nu}, \qquad\qquad
\cB_{\mu\nu}=\zeta B_{\mu\nu}-\widehat \cF_{\mu\nu},
$$
within the chiral spinor multiplet. The parameter $\zeta$ will characterize the supersymmetry deformation in 
the final lagrangian. Note for future use that the vector field \eqref{EaBab} becomes
\beq\label{EaBab2}
\cH^\mu =  {1\over2\zeta}\,\epsilon^{\mu\nu\rho\sigma}\partial_\nu \cB_{\rho\sigma}
+ {1\over2\zeta}\,\epsilon^{\mu\nu\rho\sigma}\partial_\nu(\ov\psi_\rho\gamma_\sigma\lambda)
+ {1\over4\kappa^2}\, \epsilon^{\mu\nu\rho\sigma}\ov\psi_\rho\gamma_\sigma\psi_\nu .
\eeq

Since the square of the chiral spinor multiplet $\cSp^2$ is chiral with weight $w=3$, its components can be introduced in the superconformal $F$--density action formula, $[\cSp^2]_F$. The deformed super-Maxwell 
theory is defined as
\beqn
\label{defMax}
\cL_{\text{def Maxwell}}&=& -\frac{1}{2} \Re [\cSp^2]_F
- {\vartheta\over2}\, \Im [\cSp^2]_F\,,
\eeqn
whose explicit expression is computed in \eqref{Cspsqure}.
The real part provides the super-Maxwell lagrangian with canonically-normalized kinetic terms and the
imaginary part introduces an arbitrary parameter $\vartheta$. The theory has then two parameters, $\zeta$ 
and $\vartheta$. Adding a coupling constant factor $g^{-2}$ to the first term is not necessary since it can be 
absorbed in $\cSp$ and is not observable.\footnote{We will return to this at the end of the section.} 
After some work to rearrange four-fermion terms and applying the Poincar\'e gauge-fixing
conditions \eqref{newgaugefix}, one finds
\beq
\begin{array}{rcl} \displaystyle
-\frac{1}{2} e^{-1} \Re [\cSp^2]_F &=& \displaystyle 
- {1\over4}\, \cB_{ab}\cB^{ab} - {1\over2}\, \ov\lambda\gamma^aD_a\lambda + {1\over2}\,D^2 
- {\zeta^2\over2\kappa^2}  
\crbig
&& \displaystyle
+ {1\over4}\, \ov\psi_\mu\gamma^\mu\Big({\zeta\over\kappa^2} - iD\gamma_5\Big)\lambda
- {1\over16}\,\cB_{ab}\,\ov\psi_\mu\gamma^\mu [\gamma^a,\gamma^b]\,\lambda
\crbig
&& \displaystyle
- {1\over8}\, (\ov\psi_\mu\gamma_\nu\lambda)(\ov\psi_\rho\gamma^{\rho\mu\nu}\lambda) .
\end{array}
\eeq
At this stage,
\beq
D_\mu\lambda = \partial_\mu\lambda - {3\over2}i\, A_\mu\gamma_5\lambda
+ {1\over8}\, \omega_{\mu ab} [\gamma^a,\gamma^b]\lambda
+ {1\over2\kappa^2}\psi_\mu - {i\over2}\,D\gamma_5 \psi_\mu 
+ {1\over8} [\gamma^a,\gamma^b]\psi_\mu \, \cB_{ab} .
\eeq
Defining the Poincar\'e-covariant derivative
\beq
\cD_\mu^{(P)}\lambda = \partial_\mu\lambda + {1\over8}\, \omega_{\mu ab} [\gamma^a,\gamma^b]\lambda
\eeq
with spin connection $\omega_{\mu ab}$ given by eq.~\eqref{spinconn}
leads to the deformed super-Maxwell theory
\beq
\label{realMax}
\begin{array}{rcl} \displaystyle
-\frac{1}{2} e^{-1} \Re [\cSp^2]_F &=& \displaystyle
-\frac{1}{4}\,\cB_{ab}\,\cB^{ab} - \frac{1}{2} \ov \lambda \gamma^a   \cD^{(P)} _a\lambda   
+\frac{1}{2}\, D^2 + \frac{3}{4}i\, \ov \lambda \gamma^a  \gamma_5 \lambda \,A_a
- \frac{\zeta^2}{2\kappa^4}
\crbig
&& \displaystyle
-\frac{1}{4}\, \ov\psi_c \gamma^{cab}\lambda \, \cB_{ab} 
+ \frac{\zeta}{2\kappa^2}\, \ov\psi_\mu \gamma^\mu \lambda
- {1\over8}\, (\ov\psi_\mu\gamma_\nu\lambda)(\ov\psi_\rho\gamma^{\rho\mu\nu}\lambda) .
\end{array}
\eeq
For $\zeta=0$, the lagrangian \eqref{defMax} reduces to the one of super-Maxwell theory, with also $\cB_{ab} =
- \widehat \cF_{ab}$. But for $\zeta\ne0$ the Maxwell gauge field $\cA_\mu$ is not expected to propagate 
degrees of freedom since it can be eliminated by a gauge variation of $B_{\mu\nu}$.

If the Poincar\'e spin connection $\omega_{\mu ab}(e)$ is used in expression \eqref{realMax}, 
a second four-fermion terms is generated by the contorsion tensor 
(\ref{contor}) located in the Dirac kinetic lagrangian. Using $\ov\lambda\gamma^a\lambda=0$, it reads
\beq
\label{Dirac4}
- {1\over8}\,\kappa_{abc}  \ov\lambda\gamma^{abc} \lambda
= - {1\over32}\, (\ov\lambda\gamma^{abc} \lambda)
(\ov\psi_a\gamma_b\psi_c).
\eeq

The fact that supersymmetry is broken when $C=\kappa^{-2}$ follows from the presence of a constant, 
nonlinear term in the gaugino variation:
\beq
\delta\lambda =
-  {1\over2}\left( {\zeta\over\kappa^2} - i D\gamma_5\right)\epsilon + {\rm linear},
\eeq
which also shows again that the $\zeta$-deformation cannot be absorbed or generated by a Fayet-Iliopoulos 
term inducing $\langle D\rangle\ne0$. In addition, since 
\beq
\delta \chi = 
- {2i\over\kappa^2} \gamma_5\eta  + {1\over2}\, \cH_a\gamma^a\epsilon ,
\eeq
the invariance of the Poincar\'e fixing condition $\chi=0$ implies that the Poincar\'e supersymmetry
has a parameter $\epsilon_P$ which combines $\epsilon\equiv\epsilon_P$ (${\cal Q}$ supersymmetry) with a ${\cal S}$
supersymmetry variation with parameter $\eta= {i\kappa^2\over4}\, E^a\gamma_a\gamma_5\epsilon_P$.

The $\vartheta$--term in the lagrangian \eqref{defMax} is
\beq
\label{Im5}
\begin{array}{rcl}
\displaystyle 
-{\vartheta\over2}\, \Im \cSp^2
&=& \displaystyle \vartheta\Bigl[
- {e\over8}\,\epsilon^{\mu\nu\rho\sigma} (\cB_{\mu\nu} + \ov\psi_\mu\gamma_\nu\lambda)
(\cB_{\rho\sigma} + \ov\psi_\rho\gamma_\sigma\lambda)
+ {e\over\kappa^2}\, \zeta D 
\crbig
&& \displaystyle
- {ie\over2\kappa^2}\, \zeta\, \ov\psi_\mu\gamma^\mu\gamma_5\lambda
- {i\over4}\, \partial_\mu (e\,\ov\lambda\gamma^\mu\gamma_5\lambda)\Bigr] .
\end{array}
\eeq
It depends on
\beq
\cB_{\mu\nu} + {1\over2}\,\ov\psi_\mu\gamma_\nu\lambda - {1\over2}\, \ov\psi_\nu\gamma_\mu\lambda
= \zeta B_{\mu\nu} - \partial_\mu {\cal A}_\nu + \partial_\nu {\cal A}_\mu
\eeq
and for $\zeta=0$ it reduces to the super-Maxwell expression
\beq
\label{Im6}
\begin{array}{rcl} \displaystyle
 -{\vartheta\over2}\, \Im \cSp^2_{\,{\rm Maxwell}}
&=& \displaystyle
- {e\over8}\vartheta \,\epsilon^{\mu\nu\rho\sigma} \cF_{\mu\nu}\cF_{\rho\sigma}
- {i\over4}\vartheta\, \partial_\mu (e\,\ov\lambda\gamma^\mu\gamma_5\lambda) \quad;
\quad
\cF_{\mu\nu} = \partial_\mu {\cal A}_\nu - \partial_\nu {\cal A}_\mu
\crbig
&=& \displaystyle - {\vartheta\over2}\, \partial_\mu \Bigl( e\,\epsilon^{\mu\nu\rho\sigma} 
{\cal A}_\nu\partial_\rho {\cal A}_\sigma 
+ {i\over2}\, e\,\ov\lambda\gamma^\mu\gamma_5\lambda \Bigr)\,, 
\end{array}
\eeq
which is a total derivative. 
If the theory is deformed by coupling the Maxwell theory to the compensator, {\it i.e.} if $\zeta\ne0$, it also
naturally includes a Fayet-Iliopoulos term $e\kappa^{-2} \, \vartheta\zeta \, D$ with free parameter $\vartheta$.
Since it is a derivative for $\zeta=0$, the $\vartheta$--term generates a lagrangian
with terms linear or quadratic in the linear multiplet fields $C$, $\chi$ and $B_{\mu\nu}$. Before applying the Poincar\'e gauge fixing
\eqref{newgaugefix},
\beq
\label{Im7}
\begin{array}{rcl}
\displaystyle -{1\over2}\, \Im \cSp^2
&=& \displaystyle 
- {e\over8}\,\zeta^2\,\epsilon^{\mu\nu\rho\sigma} B_{\mu\nu} B_{\rho\sigma} 
+ e\zeta\, CD - e\zeta\, \ov\lambda\chi
+ {e\over4}\zeta\,\epsilon^{\mu\nu\rho\sigma} B_{\mu\nu} \widehat \cF_{\rho\sigma} 
\crbig
&& \displaystyle
- {i\over2}e\zeta\, C \,\ov\psi_\mu\gamma^\mu\gamma_5\lambda
- {1\over4}e\zeta\, \epsilon^{\mu\nu\rho\sigma} B_{\mu\nu} \, \ov\psi_\rho \gamma_\sigma\lambda 
+ {\rm derivative}\,.
\end{array}
\eeq

Coupling the deformed super-Maxwell theory to new-minimal supergravity amounts to consider
\be
\label{defMaxnm}
\cL=\cL_{\text{new min SG}}+  \cL_{\text{def Maxwell}}= 
{3\over2} \left[L\ln{L\over S\ov S}-L\right]_D - {1\over2} \Re \Bigl[{\cSp}^2 \Bigr]_F
- {\vartheta\over2} \Im \Bigl[{\cSp}^2 \Bigr]_F\,.
\ee
In components and with Poincar\'e gauge fixing \eqref{newgaugefix}, this lagrangian reads 
\beq
\begin{array}{rcl}
e^{-1} \cL& =& \displaystyle {1\over2\kappa^2}\, R
- {1\over2\kappa^2}\, \ov\psi_\mu\gamma^{\mu\nu\rho}{\cal D}^{(P)}_\nu\psi_\rho   
\crbig 
&& \displaystyle  
-\frac{\zeta^2}{2\kappa^4} - {1\over 2}\, \ov \lambda \gamma^a \cD^{(P)} _a\lambda
- \frac{i\zeta}{2\kappa^2}\, \ov\psi_\mu \gamma^\mu\gamma_5 (\vartheta+i\gamma_5)\lambda
- {1\over8}\, (\ov\psi_\mu\gamma_\nu\lambda)(\ov\psi_\rho\gamma^{\rho\mu\nu}\lambda) 
\crbig
&& \displaystyle 
 +e^{-1} \cL_{\text{aux.} }.
\end{array}
\eeq
The auxiliary lagrangian $\cL_{\text{aux.}}$ includes the contributions of $\cB_{ab}$, of $D$ and of
the $U(1)_R$ gauge field $A_\mu$. A first expression is
\beq
\label{defaux}
\begin{array}{rcl}
{\cal L}_{\rm aux.} &=& \displaystyle 
-{3e\kappa^2\over4}\, E_\mu E^\mu 
- 3e \, A_\mu E^\mu + {3ie\over4} \,  A_\mu \, \ov\lambda\gamma^\mu\gamma_5\lambda
+ {e\over2}\, D^2 + {e\over\kappa^2}\, \zeta \vartheta D 
\crbig
&& \displaystyle
- {e\over4}\, \cB^{ab} \cB_{ab} 
- {\vartheta \over8}\,e\,\epsilon^{\mu\nu\rho\sigma} (\cB_{\mu\nu} + \ov\psi_\mu\gamma_\nu\lambda)
(\cB_{\rho\sigma} + \ov\psi_\rho\gamma_\sigma\lambda)
- {e\over4}\, \cB_{cd}  \, \ov\psi_a\gamma^{acd}\lambda,
\end{array}
\eeq
where $E^\mu$ is given in eq.~\eqref{EaBab2}. The first two terms originate from the supergravity lagrangian, 
while all others from the Maxwell lagrangian.

Working from here on with $\zeta\ne0$ and defining a new $U(1)_R$ gauge field by generalising \eqref{Aredef}
as
\beq
\label{newA}
{\zeta\over3}\,\widetilde A_\mu = A_\mu + {\kappa^2\over4} E_\mu  
+ {\kappa^2\over16}\,i \ov\lambda\gamma_\mu\gamma_5\lambda ,
\qquad\qquad\qquad 
\zeta\ne0
\eeq
leads to a second expression:
\beq
\label{defaux2}
\begin{array}{rcl}
{\cal L}_{\rm aux.} &=& \displaystyle
- e\zeta \widetilde A_\mu \Big(E^\mu - {i\over4}\,\ov\lambda\gamma^\mu\gamma_5\lambda\Big)
- {e\over4}\, \cB^{ab} \cB_{ab} 
- {e\over4}\, \cB_{cd}  \, \ov\psi_a\gamma^{acd}\lambda
\crbig
&&\displaystyle
- {\vartheta\over8}\,e\,\epsilon^{\mu\nu\rho\sigma} (\cB_{\mu\nu} + \ov\psi_\mu\gamma_\nu\lambda)
(\cB_{\rho\sigma} + \ov\psi_\rho\gamma_\sigma\lambda)
\crbig
&&\displaystyle
+ {3e\kappa^2\over64}\, (\ov\lambda\gamma^a\gamma_5\lambda)(\ov\lambda\gamma_a\gamma_5\lambda)
+ {e\over2}\, D^2 + {e\over\kappa^2}\, \zeta \vartheta D\, .
\end{array}
\eeq
The first contribution, which follows from the coupling $-3e\,A_\mu E^\mu$ of eq.~\eqref{defaux}, 
indicates that even if $\cB_{ab}$ has a two-derivative lagrangian in eq.~\eqref{defaux},
it does not propagate degrees of freedom. It is auxiliary, as in pure new-minimal supergravity, but it is not any longer a gauge 
field. It includes then six components, the number required by minimal supergravity. 
Using now expression \eqref{EaBab2} and integrating by parts, one finds an equivalent ${\cal L}_{\rm aux.}$ 
in which $\cB_{ab}$ has a simple algebraic field equation:
\beq
\label{defaux3}
\begin{array}{rcl}
 {\cal L}_{\rm aux.} 
&=&
\displaystyle
- {e\over4}\, \cB^{ab} \cB_{ab} 
- {\vartheta\over8}\,e\,\epsilon^{\mu\nu\rho\sigma} \cB_{\mu\nu} 
\cB_{\rho\sigma} 
\crbig
&&\displaystyle
- {1\over4}\,e\,\epsilon^{\mu\nu\rho\sigma} \widehat{\widetilde F}_{\mu\nu}\cB_{\rho\sigma}
- {1\over4}\,e\, \epsilon^{\mu\nu\rho\sigma}\widehat{\widetilde F}_{\mu\nu} \ov\psi_\rho\gamma_\sigma\lambda
+ {e\over2}\, D^2 + {e\over\kappa^2}\, \zeta \vartheta D 
\crbig
&&\displaystyle
+ {i\zeta\over4}\,e\, \widetilde A_\nu\,  \Bigl( {1\over\kappa^2}\,\ov\psi_\mu\gamma^{\mu\nu\rho}\gamma_5\psi_\rho
+ \ov\lambda\gamma^\nu\gamma_5\lambda \Bigr)
+ {\vartheta\over8}\,e\,\epsilon^{\mu\nu\rho\sigma} 
(\ov\psi_\mu\gamma_\nu\lambda)(\ov\psi_\rho\gamma_\sigma\lambda)
\crbig
&&\displaystyle
+ {e\over4}\, (\ov\psi_\mu\gamma^{\mu\nu\rho}\lambda)(\ov\psi_\nu\gamma_\rho\lambda)
+ {3e\kappa^2\over64}\, (\ov\lambda\gamma^a\gamma_5\lambda)(\ov\lambda\gamma_a\gamma_5\lambda).
\end{array}
\eeq
The above expression depends on the abelian field-strength of $\widetilde A _\mu $,
\beq
\widehat{\widetilde F}_{\mu\nu} = \partial_\mu  \widetilde A _\nu  - \partial_\nu  \widetilde A _\mu   
+\frac  {1}{2} \ov\psi_\mu \gamma_\nu (i\gamma_5 + \vartheta)\lambda
- \frac  {1}{2} \ov\psi_\nu \gamma_\mu (i\gamma_5 + \vartheta)\lambda.
\eeq
It is covariant under the supersymmetry variation
\beq
\delta {\widetilde A}_\mu = -{1\over2}\ov\epsilon\gamma_\mu(\vartheta + i\gamma_5)\lambda\,,
\eeq
suggesting that the supersymmetric partner of $\widetilde A_\mu$ is $\widetilde\lambda = 
(\vartheta +i\gamma_5)\lambda$.
Note that in eq.~\eqref{defaux3} 
the field strength $\widehat{\widetilde F}_{\mu\nu}$ constructed from the $U(1)_R$ gauge field 
$A_\mu$ has a $\cB\wedge \widehat{\widetilde F}$ ({\it electric--like}) interaction with the antisymmetric tensor 
while the original Maxwell field strength $\widehat{\cal F}_{\mu\nu}$ only appears in the ({\it magnetic-like}) 
combination $\cB_{\mu\nu} = \zeta B_{\mu\nu}-\widehat{\cal F}_{\mu\nu}$, which is auxiliary
(see eqs.~\eqref{realMax} and \eqref{Im5}).
As we will soon see, the emergence of the $\cB\wedge\widehat{\widetilde F}$ coupling plays a crucial role 
in the integration of the auxiliary fields and an associated electric-magnetic duality transformation
in the theory.

Eliminating $D$ leads to 
\beq
D= - {\zeta\vartheta\over\kappa^2}\,,
\eeq
implying the supersymmetry variations
\beq\label{gaugino-var}
\delta\lambda = - {\zeta\over2\kappa^2}(1 + i\vartheta\gamma_5)\epsilon + {\rm linear}\,,
\qquad\qquad
\delta\widetilde\lambda = - {\zeta\over2\kappa^2}(1+\vartheta^2)i\gamma_5\epsilon + {\rm linear},
\eeq
while the cosmological constant becomes
\beq
\Lambda= {\zeta^2\over2\kappa^4} (1+\vartheta^2).
\eeq
Finaly, eliminating $\cB_{ab}$ leads to
\beq
\label{Bis}
\cB_{ab} = - {1\over 1+\vartheta^2} \Bigl[ \vartheta\, \widehat{\widetilde F}_{ab} 
+ {1\over2}\,\epsilon_{abcd}\widehat{\widetilde F}^{cd} \Bigr]\,.
\eeq

It follows that 
\beq
\begin{array}{rcl}
{\cal L}_{\rm aux.} 
&=&
\displaystyle
- {e\over4(1+\vartheta^2)}\, \Bigl[ \widehat{\widetilde F}^{\mu\nu} \widehat{\widetilde F}_{\mu\nu} 
- {\vartheta\over2}\,\epsilon^{\mu\nu\rho\sigma} \widehat{\widetilde F}_{\mu\nu} 
\widehat{\widetilde F}_{\rho\sigma} \Bigr]
- {e\zeta^2\vartheta^2\over2\kappa^4}
- {1\over4}\,e\, \epsilon^{\mu\nu\rho\sigma} \widehat{\widetilde F}_{\mu\nu} \ov\psi_\rho\gamma_\sigma\lambda
\crbig
&&\displaystyle
+ {i\zeta\over4}\,e\, \widetilde A_\nu\,  \Bigl( {1\over\kappa^2}\,\ov\psi_\mu\gamma^{\mu\nu\rho}\gamma_5\psi_\rho
+ \ov\lambda\gamma^\nu\gamma_5\lambda \Bigr)
+ {\vartheta\over8}\,e\,\epsilon^{\mu\nu\rho\sigma} 
(\ov\psi_\mu\gamma_\nu\lambda)(\ov\psi_\rho\gamma_\sigma\lambda)
\crbig
&&\displaystyle
+ {e\over4}\, (\ov\psi_\mu\gamma^{\mu\nu\rho}\lambda)(\ov\psi_\nu\gamma_\rho\lambda)
+ {3e\kappa^2\over64}\, (\ov\lambda\gamma^\mu\gamma_5\lambda)(\ov\lambda\gamma_\mu\gamma_5\lambda)\,,
\end{array}
\eeq
so that the complete lagrangian, upon elimination of $\cB_{ab}$ and expressed in terms of $\widetilde\lambda$ and of the Poincar\'e spin connection $\omega_{\mu ab}(e)$ reads
\beq
\label{final1}
\begin{array}{rcl}
\cL& =& \displaystyle {1\over2\kappa^2}\, eR
- {e\over2\kappa^2}\, \ov\psi_\mu\gamma^{\mu\nu\rho}{\cal D}^{(P)}_\nu\psi_\rho   
-\frac{e\zeta^2}{2\kappa^4}(1+\vartheta^2) 
- \frac{ie\zeta}{2\kappa^2}\, \ov\psi_\mu \gamma^\mu\gamma_5 \widetilde\lambda
\crbig 
&& \displaystyle
- {e\over4(1+\vartheta^2)}\, \Bigl[ \widehat{\widetilde F}^{\mu\nu} \widehat{\widetilde F}_{\mu\nu} 
- {\vartheta\over2}\,\epsilon^{\mu\nu\rho\sigma} \widehat{\widetilde F}_{\mu\nu} \widehat{\widetilde F}_{\rho\sigma} 
- 2\, \ov{\widetilde\lambda} \gamma^\mu \cD^{(P)} _\mu \widetilde\lambda \Bigr]
\crbig
&& \displaystyle  
+ {ie\zeta\over4}\, \widetilde A_\nu\,  \Bigl( {1\over\kappa^2}\,\ov\psi_\mu\gamma^{\mu\nu\rho}\gamma_5\psi_\rho
+ {1\over1+\vartheta^2}\, \ov{\widetilde\lambda}\gamma^\nu\gamma_5\widetilde\lambda \Bigr)
\crbig
&&\displaystyle
- {e\over4(1+\vartheta^2)}\, \epsilon^{\mu\nu\rho\sigma} (\ov\psi_\mu\gamma_\nu
(\vartheta-i\gamma_5)\widetilde\lambda) \Big(\widehat{\widetilde F}_{\rho\sigma} - {1\over2}\,\ov\psi_\rho\gamma_\sigma\widetilde\lambda \Big)
\crbig
&&\displaystyle
+ {3e\kappa^2\over64(1+\vartheta^2)^2}\, (\ov{\widetilde\lambda}\gamma^\mu\gamma_5\widetilde\lambda)(\ov{\widetilde\lambda}\gamma_\mu\gamma_5\widetilde\lambda)
- {e\over32(1+\vartheta^2)}\, (\ov{\widetilde\lambda}\gamma^{\mu\nu\rho}\widetilde\lambda)
(\ov\psi_\mu\gamma_\nu\psi_\rho) + {\cal L}_{\rm 4,SG},
\end{array}
\eeq
with field-strength
\beq
\widehat{\widetilde F}_{\mu\nu} = \partial_\mu  \widetilde A _\nu  - \partial_\nu  \widetilde A _\mu   
+\frac  {1}{2} \ov\psi_\mu \gamma_\nu\widetilde\lambda
- \frac  {1}{2} \ov\psi_\nu \gamma_\mu \widetilde\lambda
\eeq
and with four-gravitino terms ${\cal L}_{\rm 4,SG}$ given in eq.~\eqref{4grav}.

Since 
$$
\epsilon^{\mu\nu\rho\sigma}
(\widehat{\widetilde F}_{\mu\nu}-  \ov\psi_\mu \gamma_\nu\widetilde\lambda) 
(\widehat{\widetilde F}_{\rho\sigma} - \ov\psi_\rho \gamma_\sigma\widetilde\lambda)
$$
is a derivative, one finally finds
\beq
\label{final2}
\begin{array}{rcl}
\cL& =& \displaystyle {e\over2\kappa^2}\, R
- {e\over2\kappa^2}\, \ov\psi_\mu\gamma^{\mu\nu\rho}{\cal D}^{(P)}_\nu\psi_\rho   
-\frac{e\zeta^2}{2\kappa^4}(1+\vartheta^2) 
- \frac{ie\zeta}{2\kappa^2}\, \ov\psi_\mu \gamma^\mu\gamma_5 \widetilde\lambda
\crbig 
&& \displaystyle
- {e\over4(1+\vartheta^2)}\, \Bigl[ \widehat{\widetilde F}^{\mu\nu} \widehat{\widetilde F}_{\mu\nu} 
- 2\, \ov{\widetilde\lambda} \gamma^\mu \cD^{(P)}_\mu \widetilde\lambda \Bigr]
\crbig
&& \displaystyle  
+ {ie\zeta\over4}\, \widetilde A_\nu\,  \Bigl( {1\over\kappa^2}\,\ov\psi_\mu\gamma^{\mu\nu\rho}\gamma_5\psi_\rho
+ {1\over1+\vartheta^2}\, \ov{\widetilde\lambda}\gamma^\nu\gamma_5\widetilde\lambda \Bigr)
\crbig
&&\displaystyle
+ {ie\over4(1+\vartheta^2)}\, \epsilon^{\mu\nu\rho\sigma} (\ov\psi_\mu\gamma_\nu\gamma_5\widetilde\lambda) 
\Big(\widehat{\widetilde F}_{\rho\sigma} - {1\over2}\,\ov\psi_\rho\gamma_\sigma\widetilde\lambda \Big)
\crbig
&&\displaystyle
+ {3e\kappa^2\over64(1+\vartheta^2)^2}\, (\ov{\widetilde\lambda}\gamma^\mu\gamma_5\widetilde\lambda)(\ov{\widetilde\lambda}\gamma_\mu\gamma_5\widetilde\lambda)
- {e\over32(1+\vartheta^2)}\, (\ov{\widetilde\lambda}\gamma^{\mu\nu\rho}\widetilde\lambda)
(\ov\psi_\mu\gamma_\nu\psi_\rho) 
\crbig
&& \displaystyle
+  {\cal L}_{\rm 4,SG} +\, {\rm derivative}\,,
\end{array}
\eeq
where the spin connection in the Poincare covariant derivative ${\cal D}^{(P)}_\nu$ is  $\omega_{\mu ab}(e)$ of \eqref{omegae}.

In the first line of the above expressions \eqref{final1} and \eqref{final2}, the last two terms are such that the $\zeta^2$ contributions in their supersymmetry variations
cancel. The second line is the super-Maxwell kinetic lagrangian with a gauge coupling $\sqrt{1+\vartheta^2}$. 
The third line implies that both the gravitino and 
the gaugino $\widetilde\lambda$ have $U(1)$ charge $\zeta/2$ in units of the $U(1)$ gauge coupling
of $\widetilde{A}_\mu$. 
The terms in the fourth line can also be seen in eqs.~\eqref{realMax} or \eqref{Im5}. 
Note that after canonical normalisation of the gauge and gaugino kinetic terms by a rescaling of 
$\widetilde A_\mu$ and $\widetilde\lambda$ with the factor $\sqrt{1+\vartheta^2}$, the super-Maxwell 
part of the lagrangian depends only on one parameter $q=\zeta\sqrt{1+\vartheta^2}/2$ which amounts to
the physical $U(1)$ charge of the gaugino.
In particular, the cosmological constant is given by
\beq
\label{lambda-charge}
\Lambda = {2q^2\over\kappa^4}\,.
\eeq

To exhibit electric-magnetic duality between the original $\widehat{\cal F}_{\mu\nu}$ (eq.~\eqref{cFdef})
and the dynamical $\widehat{\widetilde F}_{\mu\nu}$ (and their supersymmetry partners)
and make the connection with the global supersymmetry case of section~2, we may introduce back a 
complex gauge coupling in front of $\cSp^2$, as in eq.~\eqref{cspaction}:\,\footnote{
It could also be understood as a background value of a holomorphic function of some neutral chiral matter 
superfields of the theory. In this case, of course, additional modifications of the lagrangian 
are needed which go beyond the scope of this paper.} 
\beq
\label{cspaction2}
\cL_{\text{def Maxwell}} = -\frac{1}{2} \Im\left[\widetilde\tau \cSp^2\right]_F ,
\qquad\qquad
\widetilde\tau = \vartheta + {i\over\widetilde g\,^2}.
\eeq 
Restoring the factors of $\widetilde g$ in the above 
analysis, it is easy to see that the gauge kinetic terms in the second line of eq.~\eqref{final1} read
\beq
\label{final3}
- {e\over4}\Im{\tau} \, \Bigl[ \widehat{\widetilde F}^{\mu\nu} \widehat{\widetilde F}_{\mu\nu} 
- 2\, \ov{\widetilde\lambda} \gamma^\mu\cD^{(P)}_\mu \widetilde\lambda \Bigr]
-{1\over8}\Re{\tau}\,e\,\epsilon^{\mu\nu\rho\sigma}
\widehat{\widetilde F}_{\mu\nu} \widehat{\widetilde F}_{\rho\sigma}
\eeq
with complex coupling
\beq
\tau = -{1\over\widetilde\tau}.
\label{TtildeT}
\eeq
Hence, the physical complex coupling of the propagating super-Maxwell fields is inverted with respect
to the lagrangian coupling $\widetilde\tau$. This duality inversion is due to the presence in the first line of
the auxiliary lagrangian \eqref{defaux3} of terms quadratic in the auxiliary tensor $\cB_{ab}$, as displayed
in the solution \eqref{Bis}.
The above expression (\ref{final2}) indicates an ``electric" theory, dual to the original deformed ``magnetic" theory \eqref{defMax}, as could be expected since the starting point was the $\zeta B_{\mu\nu}-\widehat \cF_{\mu\nu}$ coupling of the original Maxwell field. 

Note also the agreement with the globally supersymmetric deformed Maxwell theory case \eqref{MagMaxwell} and \eqref{EleMaxwell}. Moreover the cosmological constant reads:
\beq
\Lambda = {\zeta^2\over2\kappa^4}\,{\ov{\tilde\tau}\tilde\tau\over\Im\tilde\tau}={\zeta^2\over2\kappa^4}\,\frac{1}{\Im\tau}\,,
\eeq
in agreement with \eqref{ccglobal} for $\xi=0$. In the next section, we show that the ``electric" supergravity theory \eqref{final2}, \eqref{final3} corresponds to a standard gauging of the $R$-symmetry with the deformation parameter $\zeta$ being the coefficient of the Fayet-Iliopoulos term.

\section{On the Fayet-Iliopoulos term in supergravity} \label{FIL}

\noindent
In the new-minimal formulation, the super-Maxwell theory is obtained from the superconformal
lagrangian
\beq
\label{LFI1}
\cL_{\text{new-min, Max}}=
{3\over2} \left[L\ln{L\over S\ov S}-L\right]_D - {1\over2} \Im \Bigl[\tau W^2 \Bigr]_F\quad;\quad \tau=\theta+\frac{i}{g^2}\,,
\eeq
where $W$ is the chiral spinor multiplet of the field-strength $\widehat \cF_{\mu\nu}$ (up to a sign), defined in \eqref{cFdef} and Appendix~\ref{AppB}. With respect to our previous discussion, $W$ is the Maxwell submultiplet 
of $\cSp$ obtained by choosing the deformation parameter $\zeta=0$. 

We have already seen in \eqref{Im6}, and it is well-known, that 
$\Im[  W^2]_F$  is a derivative irrelevant in the theory. The addition to ${\cal L}_{\text{new-min, Max}}$ of 
\beq
\label{LFI2}
{\cal L}_{\rm FI} = {3\over2}\xi\, [LV]_D,
\eeq
with a real coefficient $\xi$, generates the Fayet-Iliopoulos term. This expression couples the compensating multiplet $L$ to the real $w=0$ multiplet $V$ of the Maxwell gauge field ${\cal A}_\mu$, corresponding to the chiral field-strength spinor multiplet $W$. 
It is invariant under gauge transformation $\delta V = \Lambda+\ov\Lambda$ ($\Lambda$ is chiral with $w=0$) since
$[L(\Lambda+\ov\Lambda)]_D$ is a derivative. Notice that one can also write
\beq
\label{LFI3}
{\cal L} = \cL_{\text{new-min, Max}} + {\cal L}_{\rm FI}  =
{3\over2} \left[L\ln L - L\ln\left( \ov S e^{-\xi V} S\right) -L\right]_D - {1\over2} \Im \Bigl[\tau W^2 \Bigr]_F
\eeq
and view the Fayet-Iliopoulos term as a gauging of the $U(1)$ invariance of the (unphysical) chiral multiplet $S$.

Using the real multiplet tensor calculus, the product of the real gauge multiplet $V$ in Wess-Zumino gauge~\eqref{VecWZ}
with $L$~\eqref{newL} has components
\beq
\label{LFI4}
LV = \Big(\,\, 0, \quad 0, \quad 0, \quad C{\cal A}_a, \quad C\lambda 
- {i\over2}\,\gamma^a\gamma_5\chi {\cal A}_a, \quad
CD + E^a{\cal A}_a - \ov\chi\lambda - {1\over4}\,\ov\chi\gamma^a \gamma^b\psi_a {\cal A}_b \,\,\Big),
\eeq
where the vector field $E_a$  is given in eq.~\eqref{His1} (or in eq.~\eqref{EaBab} after Poincar\'e gauge fixing).
The residual invariance of the Wess-Zumino gauge is the bosonic gauge invariance $\delta{\cal A}_\mu = \partial_\mu \alpha$.
Inserting the $w=2$ multiplet $LV$ in the real action density formula gives
\beq
\label{LFI5}
\begin{array}{rcl}
[LV]_D &=& \displaystyle
e\,CD - e\,\ov\chi\lambda - {i\over2}e\,C\,\ov\psi_\mu\gamma^\mu\gamma_5\lambda
+ e\, {\cal A}_\mu \Bigl[ E^\mu 
+ {1\over4}\, \ov\psi_\nu[\gamma^\nu,\gamma^\mu]\chi
- {1\over4}\, C\,\epsilon^{\mu\nu\rho\sigma}\ov\psi_\nu\gamma_\rho\psi_\sigma\Bigr]
\crbig
&=& \displaystyle
{e\over2}\,\epsilon^{\mu\nu\rho\sigma} {\cal A}_\mu \partial_\nu B_{\rho\sigma}
+ e\,CD - e\,\ov\chi\lambda - {i\over2}e\,C\,\ov\psi_\mu\gamma^\mu\gamma_5\lambda.
\end{array}
\eeq
After Poincar\'e gauge fixing \eqref{newgaugefix},
\beq
\label{LFI6}
{3\over2}e^{-1}\xi[LV]_D = {3\over2}\xi\Bigl[
{1\over\kappa^2}\,D -  {i\over2\kappa^2}\,\ov\psi_\mu\gamma^\mu\gamma_5\lambda
+  {\cal A}_\mu E^\mu \Bigr]
+ {1\over2\kappa^2}\,{3\over4}i\xi{\cal A}_\nu \,\ov\psi_\mu\gamma^{\mu\nu\rho}\gamma_5\psi_\rho
\eeq
and the gravitino acquires a charge under the Maxwell $U(1)$ symmetry.

Collecting the terms in \eqref{newSG2} and \eqref{realMax} with $\zeta=0$ and combining 
them with \eqref{LFI6}, we find that the full lagrangian \eqref{LFI3} reads
\beq
\label{Ltot1}
\begin{array}{rcl}
{\cal L} &=& \displaystyle
{e\over2\kappa^2}\,[ {  R} - \ov\psi_\mu\gamma^{\mu\nu\rho}{\cal D}^{(P)}_\nu\psi_\rho ] 
-\frac{e}{4}\,\widehat \cF_{\mu\nu}\,\widehat \cF^{\mu\nu} - \frac{e}{2} \ov \lambda \gamma^a   \cD^{(P)} _a\lambda
+ \frac{e}{4}\, \ov\psi_\rho \gamma^{\mu\nu\rho}\lambda \, \widehat \cF_{\mu\nu} 
\crbig
&& \displaystyle
+ {3\over4}i\xi{\cal A}_\nu
\Bigl[ {e\over2\kappa^2}\, \,\ov\psi_\mu\gamma^{\mu\nu\rho}\gamma_5\psi_\rho
+ {e\over2}\,\ov\lambda\gamma^\nu\gamma_5\lambda \Bigr]
- {3ie\over4\kappa^2}\xi\,\ov\psi_\mu\gamma^\mu\gamma_5\lambda
\crbig
&& \displaystyle
- {e\over8}\, (\ov\psi_\mu\gamma_\nu\lambda)(\ov\psi_\rho\gamma^{\rho\mu\nu}\lambda) 
+ {3e\kappa^2\over64}\, (\ov\lambda\gamma^\mu\gamma_5\lambda)(\ov\lambda\gamma_\mu\gamma_5\lambda)
+ {\cal L}_{\rm aux.}\,,
\end{array}
\eeq
where for simplicity we rescaled the gauge coupling away, and the auxiliary field lagrangian is
\beq
\begin{array}{rcl}
{\cal L}_{\rm aux.} &=& \displaystyle
-3e \Bigl(E^\mu-{i\over4}\,\ov\lambda\gamma^\mu\gamma_5\lambda\Bigr)
\Bigl(A_\mu - {\xi\over2}\,{\cal A}_\mu + {\kappa^2\over4}\,E_\mu
+ {i\kappa^2\over16}\,\ov\lambda\gamma_\mu\gamma_5\lambda\Bigr)
\crbig
&& \displaystyle
+\frac{e}{2}\, D^2 
+ {3e\over2\kappa^2}\, \xi\,D 
-3e (\partial_\mu\phi)\Bigl(E^\mu
- {i\over4\kappa^2}\, \ov\psi_\nu\gamma^{\mu\nu\rho}\gamma_5\psi_\rho\Bigr).
\end{array}
\eeq
In the last term above, 
by following the original analysis of \cite{SW},
we introduced the Lagrange multiplier $\phi$ to impose the condition \eqref{Eaconstraint} with solution \eqref{EaBab}, so that $E^\mu$ is now an unconstrained vector. Under $U(1)_R$ variations, 
$\delta_RA_\mu = \partial_\mu\lambda_T$ and also $\delta_R\phi=-\lambda_T$, leading to
\beq
\delta_R\, {\cal L}_{\rm aux.} = {3\over2}i (\partial_\mu\lambda_T)
\Bigl[ {e\over2\kappa^2}\, \,\ov\psi_\mu\gamma^{\mu\nu\rho}\gamma_5\psi_\rho
+ {e\over2}\,\ov\lambda\gamma^\nu\gamma_5\lambda \Bigr].
\eeq
This variation cancels the $U(1)_R$ variation of the Rarita-Schwinger and Dirac kinetic lagrangians.
The theory has then local $U(1)_R$ symmetry, as expected in the new-minimal formulation, and
$A_\mu + \partial_\mu\phi$ is gauge invariant.

Redefining now the auxiliary $U(1)_R$ gauge field as
\beq
 A'_\mu = A_\mu - {\xi\over2}\,{\cal A}_\mu + {\kappa^2\over4}\,E^\mu
+ {i\kappa^2\over16}\,\ov\lambda\gamma^\mu\gamma_5\lambda
\eeq
leads to
\beq
\begin{array}{rcl}
{\cal L}_{\rm aux.} &=& \displaystyle
-3e \Bigl(E^\mu-{i\over4}\,\ov\lambda\gamma^\mu\gamma_5\lambda\Bigr)( A'_\mu + \partial_\mu\phi)
+\frac{e}{2}\, D^2 + {3e\over2\kappa^2}\, \xi\,D 
\crbig
&& \displaystyle - {3\over2}i (\partial_\mu\phi)
\Bigl[ {e\over2\kappa^2}\, \,\ov\psi_\mu\gamma^{\mu\nu\rho}\gamma_5\psi_\rho
+ {e\over2}\,\ov\lambda\gamma^\nu\gamma_5\lambda \Bigr].
\end{array}
\eeq
Solving next for $E^\mu$, $A'_\mu$ and $D$ gives 
\beq\label{auxiliary-sol}
0 = E^\mu-{i\over4}\,\ov\lambda\gamma^\mu\gamma_5\lambda =  A'_\mu + \partial_\mu\phi \,,
\qquad\quad 
D= -{3\over2\kappa^2}\, \xi \,;
\qquad\quad
{\cal L}_{\rm aux.} = - {9e\over8\kappa^4}\, \xi^2
\eeq
while $\phi$ can be eliminated by a $U(1)_R$ gauge choice.

From the second line of \eqref{Ltot1}, one deduces that the $U(1)$ charge of the fermions is $q=3\xi/4$. It follows that the final form of the lagrangian can be written as
\beq
\begin{array}{rcl}
\label{LFIfinal}
{\cal L} &=& \displaystyle
{e\over2\kappa^2}\,[ {  R} - \ov\psi_\mu\gamma^{\mu\nu\rho}{\cal D}^{(P)}_\nu\psi_\rho ] 
-\frac{e}{4}\,\widehat \cF_{\mu\nu}\,\widehat \cF^{\mu\nu} - \frac{e}{2} \ov \lambda \gamma^a   \cD^{(P)} _a\lambda
+ \frac{e}{4}\, \ov\psi_\rho \gamma^{\mu\nu\rho}\lambda \, \widehat \cF_{\mu\nu} 
\crbig
&& \displaystyle
+ iq \,{ \cA}_\nu
\Bigl[ {e\over2\kappa^2}\, \,\ov\psi_\mu\gamma^{\mu\nu\rho}\gamma_5\psi_\rho
+ {e\over2}\,\ov\lambda\gamma^\nu\gamma_5\lambda \Bigr]
- {ie\over\kappa^2}\,q\,\ov\psi_\mu\gamma^\mu\gamma_5\lambda
- {2e\over\kappa^4}\, q^2
\crbig
&& \displaystyle
- {e\over8}\, (\ov\psi_\mu\gamma_\nu\lambda)(\ov\psi_\rho\gamma^{\rho\mu\nu}\lambda) 
+ {3e\kappa^2\over64}\, (\ov\lambda\gamma^\mu\gamma_5\lambda)(\ov\lambda\gamma_\mu\gamma_5\lambda)\,,
\end{array}
\eeq
in terms of the spin connection $\omega_{\mu ab}(e,\psi)$ which still includes the contorsion tensor.\,\footnote{
See eqs.~\eqref{spinconn} and \eqref{contor}.}

The theory described by  \eqref{LFIfinal} is actually the model derived in a Poincar\'e formulation by Freedman in 1976 \cite{F}.
Indeed, it can be easily transformed from the new-minimal formulation \eqref{LFI3} to the old-minimal, exhibiting the gauging of the $R$--symmetry under which the chiral compensator becomes charged. This can be done by introducing a real multiplet $U$ with $w=2$ and a chiral $T$ with $w=0$ (so that $T+\ov T$ is a real multiplet with $w=0$). One can then rewrite \eqref{LFI3} as
\beq
\label{LFIom1}
{\cal L} = 
{3\over2} \left[U\ln U - U\ln\left( \ov S e^{-\xi V} S\right) -U -  (T+\ov T)U \right]_D - {1\over2} \Im \Bigl[\tau W^2 \Bigr]_F\, ,
\eeq
where the field equation for the multiplet $T$ indicates that $U$ is a linear multiplet: $U=L$. 
On the other hand, solving for $U$, the field equation is 
\beq
\ln\left( {U\over \ov S e^{-\xi V} S}\right) = T+\ov T
\eeq
and then
\beq
{\cal L} = - {3\over2}\, \Bigl[   \ov S e^{-\xi V} S e^{T+\ov T}  \Bigr]_D - {1\over2} \Im \Bigl[\tau W^2 \Bigr]_F\,.
\eeq
Defining the $w=1$ chiral compensator $S_0$ of the old-minimal supergravity as $S_0 = e^TS$ leads to the old-minimal formulation of the theory
\beq
\label{LFIom2}
{\cal L}_{\text{old-minimal}} = - {3\over2}\, \Bigl[ \ov S_0 e^{-\xi V} S_0 \Bigr]_D 
- {1\over2} \Im \Bigl[\tau W^2 \Bigr]_F\,,
\eeq
where the Fayet-Iliopoulos term appears as a gauging of the $U(1)_R$ symmetry acting on $S_0$.

It is now easy to check that \eqref{LFIfinal}, and thus \eqref{LFI3} or equivalently \eqref{LFIom2}, is identical to the lagrangian \eqref{final2} obtained from the deformed Maxwell theory, upon normalising the kinetic terms and expressing it in terms of the single physical parameter of the Maxwell sector which is the fermion charge 
$q = 3\xi g/4 = \zeta\sqrt{1+\vartheta^2}/2$ (upon putting back the gauge coupling in \eqref{LFIfinal} according to \eqref{LFI3}), implying the identification
\beq
\label{zetaxi}
\zeta={3\over 2}\xi\,,
\eeq 
where we restored the gauge coupling $\tilde g$ in \eqref{final2} according to \eqref{cspaction} and used eqs.~\eqref{final3} and \eqref{TtildeT}.
The equivalence of the two theories \eqref{LFI3} and \eqref{defMaxnm} with gauge coupling \eqref{cspaction2} thus suggests that the deformation in the Maxwell theory coupled to supergravity exhibits two properties:
\begin{itemize}
\item it corresponds to a ``magnetic" Fayet-Iliopoulos term of the gauged $U(1)_R$ $R$--symmetry, 
in agreement with the result in global supersymmetry discussed in section~2, since their respective gauge couplings are related by the electro-magnetic duality relation \eqref{TtildeT}; 
\item it provides a different realisation of the Freedman model on-shell (i.e. upon elimination of the auxiliary fields). 
\end{itemize}
In the next section, we establish this connection by working out the explicit form of the electric-magnetic duality in supergravity.

\section{On the electric-magnetic duality}
\label{EMdualitySUGRA}

\noindent
The fact that the deformed  and the standard Fayet-Iliopoulos theory provide different constructions of the same
Freedman model suggests that electric-magnetic duality plays a role. This is also suggested by the fact that in
both descriptions the physical Maxwell fields are not the same. 
The goal of this section is to study more precisely this role of electric-magnetic (EM) duality.

In the deformed theory of section \ref{deformedL}, before elimination of the auxiliary fields, two abelian gauge 
fields are present. The $U(1)_R$ gauge field appears explicitly (without kinetic term) and the Maxwell gauge field only appears in
$\cB_{\mu\nu}$. Since no fields have Maxwell charge, one can expect that EM duality applies
on the Maxwell field. The same situation occurs in the Fayet-Iliopoulos version of the theory described 
in section \ref{FIL}. To see this, a different form of $[LV]_D$ is useful. Starting with the second expression \eqref{LFI5} and integrating by parts leads to
\beq
\label{EM1}
\begin{array}{rcl}
[LV]_D 
&=& \displaystyle
{e\over2}\,\epsilon^{\mu\nu\rho\sigma}(\partial_\mu{\cal A}_\nu) B_{\rho\sigma}
- {i\over2}e\,C\,\ov\psi_\mu\gamma^\mu\gamma_5\lambda + e\, CD - e\, \ov\chi\lambda
\crbig
&=& \displaystyle
{e\over4}\,\epsilon^{\mu\nu\rho\sigma} \widehat \cF_{\mu\nu} B_{\rho\sigma}
- {e\over4}\,\epsilon^{\mu\nu\rho\sigma} B_{\mu\nu} \ov\psi_\rho\gamma_\sigma\lambda
- {i\over2}e\,C\,\ov\psi_\mu\gamma^\mu\gamma_5\lambda + e\, CD - e\, \ov\chi\lambda\,,
\end{array}
\eeq
up to total derivatives. After Poincar\'e gauge fixing,
\beq
\label{EM2}
[LV]_D = {e\over4}\,\epsilon^{\mu\nu\rho\sigma} \widehat \cF_{\mu\nu} B_{\rho\sigma}
- {e\over4}\,\epsilon^{\mu\nu\rho\sigma} B_{\mu\nu} \ov\psi_\rho\gamma_\sigma\lambda
- {ie\over2\kappa^2}\,\ov\psi_\mu\gamma^\mu\gamma_5\lambda + {e\over\kappa^2}\, D\,.
\eeq
Thus, the Maxwell gauge field only appears through its field-strength $\widehat \cF_{\mu\nu}$ and there is 
a $B\wedge \widehat \cF$ interaction. Applying EM duality should plausibly lead to a theory depending on 
$B - \widehat{\widetilde \cF}$, in terms of the magnetic dual $\widehat{\widetilde \cF}$ of the electric field-strength $\widehat \cF$.

Consider the product $\cSp W$ of the Maxwell $W$ with the chiral spinor superfield having 
$L$ as submultiplet. Since $\cSp W$ is chiral with $w=3$, there is a superconformal action formula which gives
\beq
\label{EM3}
\begin{array}{rcl} \displaystyle
-{1\over2}\,\Im [\cSp{ W}]_F &=& \displaystyle
{e\over8}\,\epsilon^{\mu\nu\rho\sigma}(\widehat \cF_{\mu\nu}-\ov\psi_\mu\gamma_\nu\lambda)
(\cB_{\rho\sigma}+\ov\psi_\rho\gamma_\sigma\lambda)
+ {e\over2}\, C D 
- {e\over4}\,\ov{\lambda} ( 2\chi + iC\, \gamma^\mu\gamma_5\psi_\mu)
\crbig
&=& \displaystyle
{e\over4}\,\epsilon^{\mu\nu\rho\sigma}(\partial_\mu{\cal A}_\nu)B_{\rho\sigma}
+ {e\over2}\, C D 
- {e\over4}\,\ov{\lambda} ( 2\chi + iC\, \gamma^\mu\gamma_5\psi_\mu)\,,
\end{array}
\eeq
omitting derivatives. Then, by comparing \eqref{EM1} with \eqref{EM3}, one finds
\beq
\label{EM4}
\Im [\cSp{ W}]_F = - [LV]_D + {\rm derivative}\,.
\eeq
In global Poincar\'e supersymmetry, the analogous statement is
\beq
\label{EM5}
\Dint LV = - \Fint \chi W + \Fbarint \ov{\chi W} ,
\eeq
since $W_\alpha = -{1\over4}\,\ov{DD}D_\alpha V$ and any real linear superfield can be written as
$L= D\chi-\ov D\ov\chi$.

To discuss now EM duality, we need to introduce two new multiplets, unrelated to the ones used earlier.
Firstly, a chiral spinor
$\underline\cSp$, with components $\underline\lambda$, $\underline C$, $\underline D$, 
$\underline\cB_{ab}$, $\underline\chi$. Secondly, a Maxwell multiplet $\widetilde{\cal W}$, with 
components $\widetilde\lambda$, $\widehat{\widetilde W}_{\mu\nu}$, $\widetilde D$ and gauge field 
$\widetilde W_\mu$. Consider the lagrangian contribution \eqref{EM3}: 
\beq
\label{EM6}
\begin{array}{rcl} \displaystyle
-{1\over2}\,\Im [ \underline\cSp \widetilde{\cal W}]_F &=& \displaystyle
{e\over8}\,\epsilon^{\mu\nu\rho\sigma}(\widehat{\widetilde W}_{\mu\nu}-\ov\psi_\mu\gamma_\nu\widetilde\lambda)
(\underline\cB_{\rho\sigma}+\ov\psi_\rho\gamma_\sigma\underline\lambda)
+ {e\over2}\, \underline C\widetilde D 
- {e\over4}\,\ov{\widetilde\lambda} ( 2\underline\chi + i\underline C\, \gamma^\mu\gamma_5\psi_\mu)
\crbig
&& \displaystyle + \,{\rm derivative}\,.
\end{array}
\eeq
The field equations of the components of $\widetilde{\cal W}$ imply $\underline C=\underline\chi=0$ and 
\beq
\label{EM7}
\underline\cB_{\rho\sigma} = -\partial_\rho \underline A_\sigma + \partial_\sigma\underline A_\rho 
- {1\over2}\,\ov\psi_\rho\gamma_\sigma\underline\lambda 
+ {1\over2}\, \ov\psi_\sigma\gamma_\rho\underline\lambda
\eeq
for some gauge field $\underline A_\mu$. In other words, $\widetilde{\cal W}$ variation of the action implies that $\underline\cSp=
- \underline W$, a Maxwell multiplet.
This also indicates that $\Im [\underline W\widetilde{\cal W}]_F$ is a derivative for any pair of Maxwell multiplets
and that the contribution \eqref{EM6} is invariant under the gauge transformation
\beq
\label{EM8}
\underline \cSp \qquad\longrightarrow\qquad \underline \cSp + \makebox{any Maxwell supermultiplet}\,.
\eeq

Super-Maxwell theory can then be written
as
\beq
\label{EM9}
\begin{array}{rcl}
{\cal L}_{\rm super-Maxwell} &=& \displaystyle 
-{1\over2}\Im\tau\Re [\underline\cSp^2]_F - {1\over2}\,\Re\tau \Im [\underline\cSp^2]_F 
- \Im [\underline\cSp \widetilde{\cal W}]_F
\crbig
&=& \displaystyle
-{1\over2}\Im (\tau[\underline\cSp^2]_F) - \Im [\underline\cSp \widetilde{\cal W}]_F \,,
\qquad
\tau = \theta + {i\over g^2}\,,
\end{array}
\eeq
which turns into
\beq
\label{EM10}
{\cal L}_E = -{1\over2g^2}\Re [W^2]_F + {\rm derivative}
\eeq
after the elimination of $\widetilde{\cal W}$ leading to $\underline\cSp=-W$.
On the other hand, one could instead eliminate $\underline\cSp$ using its field equation
\beq
\label{EM11}
\underline\cSp = - {1\over\tau}\,\widetilde{\cal W}.
\eeq
One then obtains another form of the theory
\beq
\label{EM12}
 {\cal L}_M = -{1\over2}\Im \Big(-{1\over\tau}[\widetilde{\cal W}^2]_F\Big)
= -{1\over2}\,{1\over g^2\tau\ov\tau}\Re [\widetilde{\cal W}^2]_F + {\rm derivative},
\eeq
which is the magnetic dual of super-Maxwell theory, with inverted complex coupling $\tilde\tau= -1/\tau$.

We next add the new-minimal supergravity \eqref{newSG} and the Fayet-Iliopoulos terms 
to the lagrangian \eqref{EM9}:\beq
\label{EM13}
{\cal L} =-{1\over2}\Im (\tau[\underline\cSp^2]_F) - \Im [\underline\cSp \widetilde{\cal W}]_F 
+ {3\over2}\xi\,\Im[\cSp\underline\cSp]_F + {3\over2} \left[L\ln{L\over S\ov S}-L\right]_D\,, 
\eeq
where $\cSp$ contains the compensator linear multiplet of the new-minimal formulation.
Again, the field equation of $\widetilde{\cal W}$ is $\underline\cSp=-W$ and the term proportional to $\xi$ becomes the Fayet-Iliopoulos contribution \eqref{LFI2}, leading to the action \eqref{LFI3}: 
\beq
\label{LFI3b}
{\cal L}_E= 
- {1\over2} \Im \Bigl[ \tau W^2 \Bigr]_F + {3\over2} \left[L\ln {L\over\ov S e^{-\xi V} S} -L\right]_D .
\eeq
On the other hand, the field equation of $\underline\cSp$ gives now
\beq
\label{EM14}
\underline\cSp = -{1\over\tau}(\widetilde{\cal W} - {3\over2}\xi\, \cSp) 
= - {1\over\tau}(\widetilde{\cal W} - \zeta\cSp),
\eeq
where we used the relation \eqref{zetaxi} between the Fayet-Iliopoulos parameter $\xi$ and the deformation parameter $\zeta$ used in section \ref{deformedL}. Replacing $\underline\cSp$ in \eqref{EM13}, one obtains the magnetic dual form of the action:
\beq
\label{EM15}
 {\cal L}_M =
- {1\over2}\,\Im 
\left [-{1\over\tau} (\widetilde{\cal W} - \zeta\cSp)^2 \right]_F
+ {3\over2} \left[L\ln{L\over S\ov S}-L\right]_D.
\eeq
The last step is to observe that, if $\zeta\ne0$ (and thus $\xi\ne0$), the gauge transformation 
\beq
\label{EM17}
\begin{array}{rcl}
\widetilde{\cal W}
\qquad&\longrightarrow&\qquad \widetilde{\cal W} + X
\crbig
\cSp \qquad&\longrightarrow&\qquad \displaystyle \cSp + {1\over\zeta} X\,,
\end{array}
\eeq
with $X$ an arbitrary Maxwell multiplet, 
leaves the compensating multiplet $L$ and thus the entire lagrangian invariant. The Maxwell multiplet $\widetilde{\cal W}$ 
can then be absorbed in the chiral spinor $\cSp$ and one can then write the magnetic lagrangian as
\beq
\label{EM18}
 {\cal L}_M=
- {1\over2}\,\Im 
\left [\tilde\tau \Upsilon^2 \right]_F
+ {3\over2} \left[L\ln{L\over S\ov S}-L\right]_D\,;\qquad \tilde\tau=-\frac{1}{\tau}\,.
\eeq
The multiplet $\Upsilon$ in this last equation, which stands for $\zeta\cSp - \widetilde{\cal W}$, reduces to a Maxwell 
multiplet if $\zeta=0$; it is the chiral spinor multiplet used  in the description of the deformed theory in section \ref{deformedL}. 
This completes the discussion of EM duality and
proves the equivalence, by EM duality, of the deformed and the Fayet-Iliopoulos version of Freedman's model. 

An alternative description of the $D$-deformation in supergravity and a derivation of the electromagnetic duality can be done 
using curved superspace techniques that we present in Appendix~C. In particular, we show that the deformation is dual to a 
magnetic Fayet-Iliopoulos term, {\it i.e.} to the Freedman model, using similar steps as in section~2 that can be generalised to 
the case of supergravity, within the framework of curved conformal superspace. 

\section{Concluding remarks}

\noindent
In summary, we have studied in this work a deformation in ${\cal N}=1$ supersymmetry transformations corresponding to a 
shift of the real $D$-auxiliary field of a Maxwell multiplet by an imaginary constant, modifying the associated supersymmetric 
Bianchi identity by an integration constant. In global supersymmetry, the deformed theory is the electric-magnetic dual of a 
theory with a Fayet-Iliopoulos term with the deformation parameter mapped to its constant coefficient. An important property 
of the deformed theory is that (electrically) charged states cannot be added, implying that magnetic monopoles cannot 
exist for a $U(1)$ with a Fayet-Iliopoulos term.

The coupling of the deformed theory to supergravity is achieved from the observation that the deformation 
can be seen as a background value of a linear multiplet which we identify with the 
linear compensating multiplet of the new-minimal formulation of ${\cal N}=1$ supergravity. 
Using the superconformal off-shell framework and introducing a general chiral spinor multiplet (contaning the 
field content of a linear and a Maxwell multiplets), we have shown that the deformed theory is again dual to a 
theory with Fayet-Iliopoulos term under electric-magnetic duality, generalising the result of global
supersymmetry. In the absence of matter, the latter is the old Freedman model (1976) which gauges the $U(1)_R$ 
$R$-symmetry, under which the gravitino and gaugino are charged and there is a positive cosmological constant 
equal to twice the square of the charge in Planck units (see eq.~\eqref{lambda-charge}). In fact, the Freedman model
can also be formulated in the new-minimal supergravity framework
in terms of a chiral spinor multiplet containing the $U(1)_R$ Maxwell multiplet together with the linear compensating multiplet. 
This formulation makes then the electro-magnetic duality between this model and the deformed theory manifest.

In the deformed theory, however, the Maxwell field is unobservable
since it becomes part of a non-dynamical two-form field component of the chiral spinor multiplet. Upon 
integrating out the latter, the auxiliary gauge field of the superconformal $R$-symmetry becomes dynamical
and the theory is shown to be on-shell, after elimination of auxiliary fields, equivalent to the Freedman model. 
This is consistent with the fact that the fermions (gravitino and gaugino) which are neutral under the orignal Maxwell field of the deformed theory lagrangian, as expected by the electric-magnetic duality, are charged under 
the $U(1)_R$ symmetry. 

The presence of a constant term in the supersymmetry variation of the gaugino, either due to the $\zeta$-deformation \eqref{gaugino-var} or due to the non-trivial expectation value of the $D$-auxiliary field in the Freedman model \eqref{auxiliary-sol}, implies the existence of a `unitary' gauge in which the gaugino vanishes and the lagrangian \eqref{final2} or \eqref{LFIfinal} simplifies considerably:
\beq
\label{Lunitary}
e^{-1}{\cal L}_{\rm unit. gauge} = \displaystyle
{1\over2\kappa^2}\,[ {  R} - \ov\psi_\mu\gamma^{\mu\nu\rho}{\cal D}^{(P)}_\nu\psi_\rho ] 
-\frac{1}{4}\, \cF_{\mu\nu}\, \cF^{\mu\nu} 
+ i{q\over 2\kappa^2} \,{ \cA}_\nu\,\ov\psi_\mu\gamma^{\mu\nu\rho}\gamma_5\psi_\rho
- {2\over\kappa^4}\, q^2\,.
\nonumber
\eeq
Due to the positive cosmological constant, the background metric is de Sitter and the gravitino describes four propagating helicities despite the absence of an explicit mass term, since local supersymmetry is completely fixed. This phenomenon is similar to the case of the Volkov-Akulov model of non-linear supersymmetry~\cite{nonlinear} coupled to supergravity for vanishing gravitino mass-term~\cite{nlsugra}, that can be obtained from the above expression by setting the gauge field $\cA_\mu.....$ to zero.

It would be interesting to generalise our analysis in the presence of matter chiral multiplets and field-dependent gauge couplings in both global and local supersymmetry. Another interesting question is to study the effects of the deformation in magnetic monopoles, which in principle can be described as charged states in the presence of a Fayet-Iliopoulos term in the dual theory. Finally, one could investigate the case of extended supergravity. In particular, the general non-trivial deformation of ${\cal N}=2$ supersymmetry transformations contains three parameters~\cite{ADM, Antoniadis:2019gbd} and is necessary for partial supersymmetry breaking~\cite{ADM, ADtM}. It would be then interesting to work out the coupling of the deformed theory to ${\cal N}=2$ supergravity,\footnote{Magnetic Fayet--Iliopoulos terms in ${\cal N}=2$
curved superspaces were described in \cite{magnetic-N=2-sugra}.}
 and the possibility of partial supersymmetry breaking even in the absence of hypermultiplets.

\section*{Acknowledgements}

\noindent
This work is supported in part by the Labex ``Institut Lagrange de Paris'' and in part by a CNRS PICS grant.
I.~A. and J.-P.~D. wish to thank the University of Queensland, Brisbane for hospitality during part of this research 
work. 
H.~J. is supported by the Swiss National Science Foundation.
 The work of G.~T.-M. was supported by the Albert Einstein Center
for Fundamental Physics, University of Bern, by the Australian Research Council (ARC)
Future Fellowship FT180100353, and by the Capacity Building Package of the University
of Queensland.

\appendix

\section{Conventions and some useful formulae} \label{AppA}

\noindent
We mostly follow the notations and conventions of ref.~\cite{FVP}, with some exceptions. All spinors are 
Majorana and for the chiral projections we use $\lambda_L=P_L \lambda$, $\lambda_R=P_R \lambda$ and $\gamma_5$ instead of $\gamma_*$. We   use the symbols $w,n$ to denote the Weyl and chiral weight, respectively. 

\begin{itemize}
\item
In local superconformal theory:  

Latin indices $a,b,c,\ldots$ denote the  Lorentz (tangent space) indices, and they
are raised and lowered by the flat metric $\eta_{ab}$, $\eta^{ab}$ = diag$(-+++)$.

Greek indices $\mu,\nu,\rho,\ldots$ are world indices, they
are lowered by the metric $g_{\mu\nu}$, raised by its inverse $g^{\mu\nu}$.

Both types of indices are related using the vierbein field $e_\mu^a$ or its inverse $e^\mu_a$. In particular, we define $\partial_a = e^\mu_a \partial_\mu$.


\item
The antisymmetric symbols $\epsilon_{abcd}$ and $\epsilon^{abcd}$ are Lorentz tensor   with  numerical values $\pm1, 0$, $\epsilon_{0123}=-\epsilon^{0123}=1$. Then,
$\epsilon^{\mu\nu\rho\sigma}$ and $\epsilon_{\mu\nu\rho\sigma}$  have Weyl weight $w=4$ and  $-4$, respectively, 
$e\epsilon^{\mu\nu\rho\sigma}$ and
$e^{-1}\epsilon_{\mu\nu\rho\sigma}$ have $w=0$ and are numerical with value $0,\pm1$.  
The matrices $\gamma^a$ are numerical while $\gamma^\mu = e^\mu_a\gamma^a$.
\end{itemize}
To eliminate the gauge fields of conformal boosts $f_\mu^a$, ${\cal S}$ supersymmetry $\phi_\mu$
and Lorentz symmetry (the spin connection $\omega_\mu{}^{ab}$), and obtain the $8_B+8_F$ Weyl 
multiplet of gauge fields, three invariant contraints are imposed on superconformal curvatures. 
They are respectively:
\beq
e^\mu_a  R_{\mu\nu}^{ab}(M) =
- {1\over2}\, \ov\psi_a\gamma_\nu R^{ab}({\cal Q}) -i \, e^{\rho b}\widetilde R_{\rho\nu}(T),
\qquad
\gamma^\mu R_{\mu\nu}({\cal Q})=0,
\qquad
R_{\mu\nu}^a (P)=0
\eeq
and for the gauge field of Weyl symmetry, $b_\mu=0$ as part of the gauge-fixing to Poincar\'e symmetry. 
The third constraint leads then to eq.~\eqref{spinconn} while the first and second lead to the useful formula 
\beq
\begin{array}{rcl}
e^\mu_a f_\mu^a &=& \displaystyle -{1\over 12}\, \Bigl[ R
+ {1\over2}\,\ov\psi_\mu\gamma^{\mu\nu\rho}{\cal D}^{(P)}_\nu\psi_\rho
- {3i\over4}\,\ov\psi_\mu\gamma^{\mu\nu\rho}A_\nu\gamma_5\psi_\rho \Bigr],
\crbig
\gamma^\mu\phi_\mu &=& \displaystyle - {1\over12}\, [\gamma^\mu,\gamma^\nu]{\cal R}_{\mu\nu}(\psi),
\qquad\qquad
[\gamma^\mu,\gamma^\nu]\phi_\nu \,\,=\,\,  - {1\over2}\, \gamma^{\mu\nu\rho} \, {\cal R}_{\nu\rho}(\psi)
\end{array}
\eeq
where
\beq
{\cal R}_{\mu\nu}(\psi) = {\cal D}^{(P)}_\mu\psi_\nu - {\cal D}^{(P)}_\nu\psi_\mu
-{3\over2}iA_\mu \gamma_5\psi_\nu  + {3\over2}iA_\nu \gamma_5 \psi_\mu.
\eeq

For a chiral multiplet  $\mathcal Z=(Z, \chi, F)$ with weight  $w=n=3$, the invariant $F$--density is given by  
\be\label{FtermFormla}
e^{-1} [\mathcal Z]_F=   F+\frac{1}{\sqrt 2}\,\ov\psi_\mu \gamma^\mu \chi_L 
+\frac12\, Z\, \ov\psi_\mu \gamma^{\mu\nu} \psi_{\nu R},
\qquad\qquad
\gamma^{\mu\nu} = {1\over2}[\gamma^\mu,\gamma^\nu].
\ee
For a real multiplet  $V=(C, \chi,H,B_a,\lambda, D)$ with   weight  $w=2$, $n=0$, the $D$--density is given 
by
\beqn\label{DtermFormla}
e^{-1} [V]_D &=&D+ \Box C - {i\over2}\, \ov\psi_\mu\gamma^\mu\gamma_5(\lambda+\gamma^b D_b \chi)
- {1\over4}\, \ov\psi_\mu \gamma^{\mu\nu} (H\psi_{\nu L} +\ov H\psi_{\nu R}),
\eeqn
where the superconformal d'alembertian is 
$\Box = D^a D_a = e^{\mu a}D_\mu D_a$, $D_a=e^\nu_a D_\nu$, and we use $D_\mu$ for the covariant derivative of a given field with respect to the relevant local symmetry. The curvature constraints are used to 
prove the invariance of these densities. 

Some rules of tensor calculus can be found in  \cite{FVP, KU}. 

Useful identities for $\gamma$-matrices include:
\beq
\begin{array}{ll}
\gamma^{ab}\gamma_5 =\frac{i}{2} \epsilon^{abcd}\gamma_{cd}, \qquad &
\gamma^{abc}\gamma_5 = i \epsilon^{abcd}\gamma_{d},
\crbig
\gamma_a \gamma_{bc} =\gamma_{abc} +\eta_{ab}\gamma_c  -\eta_{ac}\gamma_b, \qquad &
\gamma_{ab} \gamma_c =\gamma_{abc} +\eta_{bc}\gamma_a  -\eta_{ac}\gamma_b.
\end{array}
\eeq
Two useful four-fermion identities are
\beq
\label{4fermion}
(\ov \lambda \lambda_L)(\ov \psi_\mu [\gamma^\mu, \gamma^\nu] \psi_{\nu R} )+c.c.
=2(\ov\psi_\mu \gamma_\nu \lambda)(\ov\psi_\rho \gamma^{\mu\nu\rho} \lambda)
= 2(\ov\psi_\mu \gamma_\nu \gamma_5 \lambda)(\ov\psi_\rho \gamma^{\mu\nu\rho}\gamma_5 \lambda)
\eeq
and
\beqn\label{4fermion2}
(\bar \lambda \lambda_L)(\bar \psi_\mu [\gamma^\mu, \gamma^\nu] \psi_{\nu R} )-c.c. 
= 2i   \epsilon^{\mu\nu\rho\sigma}(\bar\psi_\mu \gamma_\nu \lambda)(\bar\psi_\rho \gamma_{\sigma}  \lambda)
+2 (\ov \psi_\nu \gamma^\nu \psi_\mu)(\ov\lambda \gamma^\mu \gamma_5 \lambda).
\eeqn

\section{The superconformal chiral spinor multiplet} \label{AppB}

\subsection{Chiral spinor multiplet in the real field basis and its decomposition}

\noindent
The lowest component of the chiral spinor multiplet is a spinor $\lambda$ with weights $w=n=3/2$.
Its highest component is a second spinor $\chi$ with $w=5/2$ and $n=-3/2$. The eight bosonic fields are 
at the intermediate level with $w=2$ and $n=0$. They form a Lorentz chiral bispinor, four complex fields
in Lorentz representation $(\bf 2,\bf 1)\times(\bf 2,\bf 1) = (\bf 1+\bf 3,\bf 1)$. They admit an equivalent formulation in terms of an 
antisymmetric tensor $\cB_{ab}$ and two real scalars $C$ and $D$. Since a detailed discussion of this 
superconformal multiplet does not seem to be available in the literature, 
this Appendix provides the necessary information.

In this field basis, the ${\cal Q}$ (parameter $\epsilon$) and ${\cal S}$ (parameter $\eta$) supersymmetry 
variations are\,\footnote{In the conventions of ref.~\cite{FVP}.}
\beq
\label{chimult}
\begin{array}{rcl}
\delta\lambda &=& \displaystyle
- {1\over2}\, (C-iD\gamma_5) \epsilon 
- {1\over8} [\gamma^a,\gamma^b]\epsilon \, \cB_{ab} ,
\crbig
\delta\, C &=& \displaystyle {i\over2} \, \ov\epsilon \gamma_5\chi, 
\crbig
\delta\, D &=& \displaystyle 
{i\over2}\ov\epsilon\gamma_5\gamma^a D_a\lambda
-{1\over2}\ov\epsilon\chi,
\crbig
\delta \cB_{ab} &=& \displaystyle -{i\over4}\, \ov\epsilon[\gamma_a,\gamma_b]\gamma_5\chi
-{1\over2}\,\ov\epsilon\gamma_a D_b\lambda 
+{1\over2}\,\ov\epsilon\gamma_bD_a\lambda - {1\over2}\, \ov\eta[\gamma_a,\gamma_b]\lambda,
\crbig
\delta\, \chi &=& \displaystyle- {i\over2}\,\gamma_5\gamma^a\epsilon\,D_a C 
- 2i\gamma_5\eta C + {1\over2}\, \cH_a\gamma^a\epsilon .
\end{array}
\eeq
In the last variation,
\beq\label{EDB}
\cH^d = {1\over2}\,\epsilon^{dabc} D_a \cB_{bc}
\eeq
with variation
\beq
\delta \cH^d = -{1\over4}\,\ov\epsilon[\gamma^d,\gamma^a] D_a\chi - {3\over2}\,\ov\eta\gamma^d\chi.
\eeq
The covariant derivatives are
\beq
\label{Appcovder}
\begin{array}{rcl}
D_\mu C &=& \displaystyle\partial_\mu C - 2\,b_\mu C - {i\over2}\, \ov\psi_\mu\gamma_5\chi,
\crbig
D_\mu\lambda &=& \displaystyle \partial_\mu\lambda
-{3\over2}(b_\mu+ i A_\mu\gamma_5)\lambda 
+ {1\over8}\omega_{\mu ab}[\gamma^a,\gamma^b]\lambda 
\crbig
&& \displaystyle
+ {1\over2}\, C \psi_\mu - {i\over2}\,D\gamma_5 \psi_\mu
+ {1\over8} [\gamma^a,\gamma^b]\psi_\mu \, \cB_{ab} ,
\crbig
D_\mu \cB_{bc} &=& \displaystyle \partial_\mu \cB_{bc} - 2\,b_\mu \cB_{bc} - \omega_{\mu b}{}^d \cB_{dc}
- \omega_{\mu c}{}^d \cB_{bd}
\crbig
&& \displaystyle
+ {i\over4}\, \ov\psi_\mu[\gamma_b,\gamma_c]\gamma_5\chi
+{1\over2}\,\ov\psi_\mu\gamma_b D_c\lambda 
-{1\over2}\,\ov\psi_\mu\gamma_c D_b\lambda + {1\over2}\, \ov\phi_\mu[\gamma_b,\gamma_c]\lambda,
\crbig
D_\mu \chi &=& \displaystyle \partial_\mu \chi - {5\over2}\, b_\mu\chi + {3\over2}iA_\mu\gamma_5\chi
+ {1\over8}\, \omega_{\mu ab}[\gamma^a,\gamma^b]\chi 
\crbig
&& \displaystyle
+ {i\over2}\,\gamma_5\gamma^a\psi_\mu \,D_a C 
+ 2i\gamma_5\phi_\mu C - {1\over2}\, \cH_a\gamma^a\psi_\mu. 
\end{array}
\eeq
The $8_B+8_F$ chiral spinor multiplet has two submultiplets with $4_B+4_F$ fields.

\paragraph{Maxwell multiplet.} Firstly,
choosing $C=\chi=0$ also requires $\cH^a=0$ and $\cB_{ab}$ verifies then Bianchi identity $D_{[a}\cB_{bc]}=0$.
The fields $\lambda$, $D$ and $\cB_{ab}= - \widehat \cF_{ab}$ form a Maxwell multiplet with
\beq
\widehat \cF_{\mu\nu} = \partial_\mu \cA_\nu - \partial_\nu \cA_\mu 
+ {1\over2}\, \ov\psi_\mu\gamma_\nu\lambda
- {1\over2}\, \ov\psi_\nu\gamma_\mu\lambda
\eeq
which is indeed the covariant field-strength of a gauge field. 

The supersymmetry variations read
\beq\label{susyTsfmaxwell}
\begin{array}{rcl}
\delta \cA_\mu &=& \displaystyle -{1\over2}\, \ov\epsilon\gamma_\mu\lambda ,
\crbig
\delta \widehat \cF_{ab} &=&  \displaystyle {1\over2}\, \ov\epsilon\gamma_a D_b\lambda
- {1\over2}\, \ov\epsilon\gamma_b D_a\lambda
+ {1\over2}\, \ov\eta [\gamma_a,\gamma_b] \lambda ,
\crbig
\delta\, \lambda &=& \displaystyle {1\over8}\, [\gamma^a,\gamma^b]\epsilon \, \widehat \cF_{ab}
+ {i\over2}\,\gamma_5\epsilon\, D,
\crbig
\delta\, D &=& \displaystyle {i\over2}\, \ov\epsilon\gamma_5\gamma^\mu D_\mu\lambda .
\end{array}
\eeq
The covariant derivative in these expressions is 
\beq
\begin{array}{rcl}
D_\mu\lambda &=& \displaystyle \partial_\mu\lambda
-{3\over2}(b_\mu+ i A_\mu\gamma_5)\lambda 
+ {1\over8}\omega_{\mu ab}[\gamma^a,\gamma^b]\lambda 
- {1\over8}\, [\gamma^a,\gamma^b]\psi_\mu\widehat \cF_{ab}
- {i\over2}\,D\gamma_5 \psi_\mu
\crbig
&\equiv& D_\mu\lambda|_{Maxwell}.
\end{array}
\eeq

Returning to the components of the chiral spinor multiplet, 
another expression for $\cH^\mu = e^\mu_d \cH^d$ is
\beq
\begin{array}{rcl}
\cH^\mu &=& \displaystyle {1\over2}\, \epsilon^{\mu\nu\rho\sigma}\partial_\nu \cB_{\rho\sigma}
- {1\over4}\, \ov\psi_\nu[\gamma^\nu,\gamma^\mu]\chi + {1\over4}\, \epsilon^{\mu\nu\rho\sigma}
\ov\psi_\nu\gamma_\rho\psi_\sigma C
\crbig
&& \displaystyle
+ {1\over2}\,\epsilon^{\mu\nu\rho\sigma}\ov\psi_\nu\gamma_\rho D_\sigma\lambda|_{Maxwell}
+ {1\over4}\,\epsilon^{\mu\nu\rho\sigma}\ov\phi_\nu[\gamma_\rho,\gamma_\sigma]\lambda
- {1\over4}\epsilon^{\mu\nu\rho\sigma} \ov\psi_\nu\gamma^\kappa\psi_\rho \, \widehat \cF_{\sigma\kappa}.
\end{array}
\eeq
In a Maxwell multiplet, the Bianchi identity leads to
\beq
{1\over2}\,\epsilon^{\mu\nu\rho\sigma}
\, \partial_\nu\widehat \cF_{\rho\sigma}
= \epsilon^{\mu\nu\rho\sigma} \Bigl(
{1\over2}\,\ov\psi_\nu\gamma_\rho D_\sigma\lambda|_{Maxwell}
+ {1\over4}\,\ov\phi_\nu [\gamma_\rho,\gamma_\sigma] \lambda
- {1\over4}\, \ov\psi_\nu\gamma^a\psi_\rho \,\widehat \cF_{\sigma a} \Bigr)
\eeq
which implies
\beq
\label{His1}
\begin{array}{rcl}
\cH^\mu &=& \displaystyle 
{1\over2}\, \epsilon^{\mu\nu\rho\sigma}\partial_\nu B_{\rho\sigma}
- {1\over4}\, \ov\psi_\nu[\gamma^\nu,\gamma^\mu]\chi + {1\over4}\, \epsilon^{\mu\nu\rho\sigma}
\ov\psi_\nu\gamma_\rho\psi_\sigma C,
\crbig
B_{\mu\nu} &=& \cB_{\mu\nu} + \widehat \cF_{\mu\nu}.
\end{array}
\eeq

The Maxwell multiplet can be alternatively obtained from the real vector multiplet  $V = (C, \chi,H,B_a, \lambda, D)$ with weight $w=n=0$. The gauge variation of $V$ is 
\beq
\delta_g\, V ={\cal V}({\cal Z}+\ov{\cal Z})
\eeq
where $\mathcal Z=(Z,\varsigma , F)$ is a $w=0$ chiral multiplet and ${\cal V}({\cal Z}+\ov{\cal Z})$ is
the embedding of ${\cal Z}+\ov{\cal Z}$ in a vector multiplet with $w=0$.
In components, the gauge variations read  
\beq\label{VecGaugeTsf}
\begin{array}{l}
\delta_g\, C= Z+\ov Z, \qquad
\delta_g\, \chi= -\sqrt2i\gamma_5\varsigma ,
\qquad
\delta_g\, H= - 2\,F,
\crbig \displaystyle
\delta_g\, B_a = i\,D_a(Z-\ov Z) = i \partial_a(Z-\ov Z) - {i\over\sqrt2}\, \ov\psi_a \gamma_5\varsigma 
= i \partial_a(Z-\ov Z) + {1\over2}\, \ov\psi_a \delta_g\chi,
\crbig
\delta_g\, \lambda = \delta_g\, D = 0. 
\end{array}
\eeq 
We can further define
\beq
\cA_a = e_a^\mu \cA_\mu = B_a - {1\over2}\, \ov\psi_a \chi
\eeq
which has the gauge variation $\delta_g\,\cA_\mu = i\partial_\mu(Z-\ov Z)$ 
required for a Maxwell gauge field. 
The supersymmetry variation of $\cA_\mu$ is given by 
\beq
\delta \cA_\mu = {1\over2}\,\ov\epsilon\gamma_\mu\lambda ,
\eeq
as expected for the gauge field of a Maxwell multiplet in Wess-Zumino gauge with $\lambda$ and $D$ 
partner components in \eqref{susyTsfmaxwell}.

The  gauge transformation in \eqref{VecGaugeTsf} enables us to impose the Wess-Zumino gauge condition on the vector 
multiplet:
\beq\label{VecWZ}
V_{\text{WZ}} = \Bigl(  0,\quad 0,\quad 0, \quad \cA_a + i \partial_a(Z-\ov Z), \quad \lambda, \quad D
\quad\Bigr).
\eeq
The supersymmetry variations of $\cA_\mu$, $\lambda$ and $D$ are those of a Maxwell multiplet \eqref{susyTsfmaxwell}. 

\paragraph{Linear multiplet.}
The second submultiplet is real and linear. In the chiral spinor multiplet, fields $C$, $\chi$ and $\cH^a$ transform into each others and they represent the algebra (the double variations close) if $D^a\cH_a=0$. Variations of the
real linear multiplet are \cite{dWR}
\beq
\label{linvar}
\begin{array}{rcl}
\delta\, C &=& \displaystyle {i\over2} \, \ov\epsilon \gamma_5\chi, 
\crbig
\delta\, \chi &=& \displaystyle- {i\over2}\,\gamma_5\gamma^a\epsilon\,D_a C 
- 2i\gamma_5\eta C + {1\over2}\, E_a\gamma^a\epsilon ,
\crbig
\delta E^a &=& \displaystyle -{1\over4}\,\ov\epsilon[\gamma^a,\gamma^b]D_b\chi - {3\over2}\,\ov\eta\gamma^a\chi
\end{array}
\eeq
with covariant derivatives
\beq
\begin{array}{rcl}
D_\mu C &=& \displaystyle \partial_\mu C - 2 \,b_\mu C - {i\over2}\,\ov\psi_\mu\gamma_5\chi,
\crbig
D_\mu \chi &=& \displaystyle \partial_\mu \chi - {5\over2}\, b_\mu\chi + {3\over2}iA_\mu\gamma_5\chi
+ {1\over8}\, \omega_{\mu ab}[\gamma^a,\gamma^b]\chi 
\crbig
&& \displaystyle
+ {i\over2}\,\gamma_5\gamma^a\psi_\mu \,D_a C 
+ 2i\gamma_5\phi_\mu C - {1\over2}\, \cH_a\gamma^a\psi_\mu
\end{array}
\eeq
and since one can rewrite $D^a\cH_a=0$ as  
\beq 
\label{DH7}
0= \displaystyle \partial_\mu \Bigl( e \cH^\mu + {e\over4}\, \ov\psi_\nu[\gamma^\nu,\gamma^\mu]\chi
- {e\over4}\, \epsilon^{\mu\nu\rho\sigma} \ov\psi_\rho\gamma_\sigma\psi_\nu \, C \Bigr),
\eeq
the solution is actually eq.~(\ref{His1}) \cite{dWR, KU}. 

The linear multiplet can be embedded in a real multiplet with weights $w=2,n=0$ as follows:
\be\label{LinearReal}
L= \Bigl( C,\quad \chi,\quad  H=0,  \quad B_a=-\cH_a, \quad \lambda=-\gamma^b D_b \chi, \quad D=-\Box C
\Bigr) , \qquad D^a \cH_a=0.
\ee
Note that $[L]_D=0$ up to derivative.

We can in principle decompose the chiral spinor multiplet into two submultiplets: a Maxwell multiplet with fields
$ \lambda$, $D$, $\widehat \cF_{ab}$ and a real linear multiplet with field  $\chi$, $C$, $B_{ab}$ with 
\be
\cB_{ab} =   B_{ab}-\widehat \cF_{ab}.
\ee
This decomposition is unstable under supersymmetry, but it is consistent. There are apparently 
two $E^a$'s: in the chiral 
spinor multiplet $E_a$ is defined as the covariant field-strength of $\cB_{ab}$ in eq.~\eqref{EDB}, 
while in the linear 
submultiplet, $E_a$ is defined by the constraint $D^a E_a =0$, eq.~\eqref{DH7}. The two definitions are 
consistent since they lead to the same solution \eqref{His1} which only depends on the fields of the linear 
submultiplet. 

\subsection{Chiral spinor multiplet in chiral multiplet basis}

\noindent
Since the chiral spinor multiplet is chiral with $w=n=3/2$, the fields $\lambda$, $C$, $D$, $\cB_{ab}$ and $\chi$ 
can be alternatively written as fields of a chiral multiplet, with components ({\it spinor}, {\it bispinor}, {\it spinor}):
\beq
\begin{array}{rl} 
w=n=3/2:\qquad& \lambda_L,
\crbig
w=2, n=0:\qquad& \displaystyle -{1\over\sqrt2}\left[ (C-iD) \mathbb{I} 
+ {1\over8}\, (\cB_{ab} + {i\over2}\,\epsilon_{abcd}\cB^{cd})[\gamma^a,\gamma^b] \right]P_L ,
\crbig
w=5/2, n=-3/2:\qquad& -2i\,\chi_L - \gamma^a\cD_a\lambda_R .
\end{array}
\eeq
These fields transform as expected for the components of a chiral multiplet with $w=n=3/2$. 
This basis is especially useful for computing the square of the chiral spinor multiplet which has 
$w=n=3$ and can then be used to obtain a superconformal $F$--density action formula.  

The square $\cSp^2 =(Z,\varsigma , F)$ of the chiral spinor multiplet has components
\beq
\begin{array}{ll}
w=n=3: &
Z=\ov\lambda\lambda_L  ,
\crbig 
w=\frac7 2, \,\, n=\frac 3 2:   \quad &
\varsigma =  -\sqrt2\,(C-iD) \lambda_L + {\sqrt2\over4}\, \cB_{ab}[\gamma^a,\gamma^b]\,\lambda_L,
\crbig 
 w=4, \,\,n=0: & F= (C-iD)^2
 - {i\over4}\,\epsilon^{abcd} \cB_{ab}\cB_{cd} 
+ {1\over2}\, \cB_{ab}\cB^{ab} + 2\, \ov\lambda\gamma^a D_a\lambda_R
+ 4i\,\ov\lambda\chi_L,
\end{array}
\eeq
where the covariant derivative is given in \eqref{Appcovder}.
This $w=n=3$ chiral multiplet leads to the superconformal $F$--density formula
\beq
\label{Cspsqure}
\begin{array}{rcl}
e^{-1}\,[\Upsilon^2]_F &=& \displaystyle 
(C-iD)^2 + {i\over4}\,\epsilon^{abcd} \cB_{ab}\cB_{cd} 
+ {1\over2}\, \cB_{ab}\cB^{ab} + 2\, \ov\lambda\gamma^aD_a\lambda_R
+ 4i\,\ov\lambda\chi_L
\crbig
&& \displaystyle
- (C-iD) \, \ov\psi_\mu\gamma^\mu\lambda_L
+ {1\over4}\,\cB_{ab}\,\ov\psi_\mu\gamma^\mu [\gamma^a,\gamma^b]\,\lambda_L 
\crbig
&& \displaystyle
+ {1\over4}\, (\ov\lambda\lambda_L)(\ov\psi_\mu[\gamma^\mu,\gamma^\nu]\psi_{\nu R}).
\end{array}
\eeq

Then we can further compute its real part 
\beqn\label{cSp2FRe}
 e^{-1} \Real [\cSp^2 ]_F&=&
\frac12\cB_{ab}\,\cB^{ab}- \frac{3i}{2} \ov \lambda \gamma^a  \gamma_5 \lambda \,A_a
+  \frac12   \ov\psi_c \gamma^{cab}\lambda \, \cB_{ab}
+ C^2- D^2
- C \ov\psi_\mu \gamma^\mu \lambda     
    \nonumber\\& &
      +  \ov \lambda \gamma^a   \cD^{(P)} _a\lambda   
   +\frac18\Big(
(  \ov \lambda \lambda_L)(\ov \psi_\mu [\gamma^\mu, \gamma^\nu] \psi_{\nu R} )+c.c. 
\Big)
+2i \, \ov \lambda \gamma_5 \chi
\qquad\qquad
\eeqn
  and imaginary part 
\beqn\label{cSp2FImag}
 e^{-1} \Imag [\cSp^2 ]_F&=&
-2 CD+ {1\over4}\,\epsilon^{abcd} \Big( \cB_{ab} + \ov \psi_a \gamma_b \lambda   \Big)
  \Big(\cB_{cd}+\ov \psi_c \gamma_d \lambda   \Big)
\nonumber\\& &
+i C\ov\lambda \gamma^\mu \gamma_5 \psi_\mu
 +2  \ov\lambda\chi 
+\frac i 2 e^{-1} \p_\mu (   e e_a^\mu \ov\lambda\gamma^a \gamma_5 \lambda  )\,,
\eeqn
where we used the four-fermion identity~\eqref{4fermion2}, as well as the following relation
\footnote{
A useful equation to show \eqref{fermionspin}   is 
$
 \p_\mu (e e^\mu{}_a ) + 3 eb_a +\omega_{\mu a b}  e^{\mu b}
 +  \frac12 e e^\rho{}_a  \ov \psi_\mu \gamma^\mu \psi_\rho=0
$ using the curvature constraint $R_{\mu\nu}^a(P)=0$.
} 
\be\label{fermionspin}
2 e\ov\lambda\gamma^\mu \gamma_5 \cD^{(P)}_\mu \lambda 
= \p_\mu (   e e_a^\mu \ov\lambda\gamma^a \gamma_5 \lambda  )
 +  \frac12 e e^\rho{}_a  (\ov \psi_\mu \gamma^\mu \psi_\rho )( \ov\lambda  \gamma^a\gamma_5 \lambda)
 +3e \ov\lambda\gamma^a \gamma_5 \lambda \, b_a\,.
\ee

\section{Deformed Maxwell theory in curved superspace}
\label{superspace}

\noindent
In this Appendix we describe a deformed vector multiplet
in curved superspace. 
To make contact with the superconformal tensor calculus,
it is natural to start by employing the \emph{conformal superspace} approach to four-dimensional $\cN=1$ 
conformal supergravity developed by Butter in \cite{Butter}
(see also the seminal work \cite{Kugo:1983mv}).
In this formalism, the superconformal group $SU(2,2|1)$ is manifestly gauged  in a curved superspace 
with covariant derivatives
\beq
{\nabla}_A = (\nabla_a , \nabla_\a,\ov\de^\ad)=E_A{}^M \Big(
\pa_M
- {\mathbf h}_M{}^{\underline{I}}\cM_{\underline{I}}
\Big)
~,
\eeq
where $E_A{}^M=E_A{}^M(x,\q,\qb)$ is the superspace inverse vielbein\footnote{Which includes the gauge fields 
$e_\mu^a (x)$ of space-time translations and the gravitino $\psi_\mu(x)$ of ${\cal Q}$--supersymmetry. } 
while ${\mathbf h}_M{}^{\underline{I}}={\mathbf h}_M{}^{\underline{I}}(x,\q,\qb)$
are gauge connections for all the superconformal generators except for translations and 
${\cal Q}$-supersymmetry:
$\cM_{\underline{I}}=(M_{ab} , \mathbb A , \mathbb D , K_a, {\cal S}_\a, \ov{\cal S}^\ad )$.\footnote{The notation in this
Appendix differs from the rest of the paper and it adheres (up to some changes in nomenclature) 
to the one of \cite{Butter,Binetruy:2000zx}, which is largely based on \cite{WB}.
For example, we decompose four-dimensional Majorana spinors
in chiral and anti-chiral parts. Compared to the flat superspace of section
\ref{global-susy}, following  \cite{Butter,Binetruy:2000zx}, the spinor covariant derivatives
satisfy the conjugation rule
$\deb_\ad=\overline{(\de_\a)}$.
Moreover, 
the normalisation of the $U(1)_R$ generator
$\mathbb A$ is 2/3 of the generator $T$ used in \cite{FVP} and earlier in this paper. 
Chiral weights in the two notations are related by $w_{\mathbb A}=2/3 \,n$
and the spinor covariant derivatives satisfy
$[\mathbb A,\de_\a]=-i \de_\a$, $[\mathbb A,\deb^\ad]=i \deb^\ad$. } 
We refer the reader to \cite{WB,Binetruy:2000zx,superspace-reviews} for reviews on
supergravity in superspace while we refer to \cite{Butter} for detail on conformal superspace
that we will assume in this Appendix. 
Note that the off-shell $8_B+8_F$ Weyl multiplet
and the transformations of superconformal multiplets, can 
be derived following a $\q=\qb=0$ component-field projection  
(see refs.~\cite{Butter,Kugo:2016zzf} for more detail).

An abelian vector multiplet coupled to conformal supergravity 
is described by a superfield $\bW_\a$ field strength which is  a superconformal chiral ($\deb^\ad \bW_\a=0$) 
of weights $(3/2,1)$ satisfying the Bianchi identity
\bea
\de^\a\mathbf{W}_\a=\deb_\ad\ov{\mathbf{W}}^\ad
~.
\label{undefBianchiW}
\eea
This is formally identical to the flat superspace one, eq.~\eqref{Bianchi}, as well as
its solution which reads
($\de^2:=\de^\a\de_\a$, $\deb^2=\deb_\ad\deb^\ad$)
\bea
\bW_\a:=-\frac{1}{4}\deb^2\de_\a\bV
~,\qquad
\ov\bW^\ad:=-\frac{1}{4}\de^2\deb^\ad\bV
~,
\label{vector-sol}
\eea
where  $\bV$ is a real scalar of weights $(0,0)$ and gauge transformation
$\d_g\bV=(\mathbf\Lambda+\ov{\mathbf\Lambda})$, $\deb^\ad\mathbf\Lambda=0$.
The Maxwell theory's action is
based on the (anti-)chiral locally superconformal action principle (equivalent to the 
tensor calculus $F$-term density formula \cite{Butter})
\beq
S_{\text{Max}}=
- {1\over2} \Im \Bigl[\tau
\int d^4x\, d^2\theta\, \cE\, \bW^2 \Bigr], \qquad\qquad \tau=\theta+\frac{i}{g^2}.
\eeq

Next we want to analyse the possible deformation of the vector multiplet Bianchi identity in superspace.
Note that, thanks to the algebra of $\de_A$,
which for instance implies $\de_\a\de_\b\de_\g\equiv0$ as in the flat superspace \cite{Butter},
equalities $\de^2(\de^\a\bW_\a)=0$ and $\deb^2(\de^\a\bW_\a)=0$ hold. These properties rely 
on the weights of $\bW_\a$ and $\de^\alpha\bW_\a$.
This implies that the following deformation of the Bianchi identity 
is consistent:\footnote{Note that this constraint arises as the obstruction
of the closure of the super two-form associated with an abelian vector multiplet
induced by the closed super three-form of a linear multiplet compensator.
This fits with the description of the abelian tensor hierarchy for $4D$, $\cN=1$
supersymmetry  \cite{Gates:1980ay,Yokokura:2016xcf}.
}
\be
\de^\a\bW^{\rm def}_\a-\deb_\ad\ov\bW^{\rm def}{}^\ad=-4i\zeta \, \widehat{\bL}\,,
\label{defWBI}
\ee
where $\widehat{\bL}$ is a real linear superfield of weights $(2,0)$ satisfying by definition
\be
\deb^2\widehat{\bL}=\de^2\widehat{\bL}=0\,.
\label{de2L}
\ee
We also define $\widehat{\bL}$ as the compensating superconformal multiplet leading to new-minimal Poincar\'e supergravity.
Then eq.~\eqref{defWBI} provides the curved superspace interpretation of the constraint \eqref{localcSpL}
(the factor of $i$ is due to the different convention used in this Appendix).

A solution of \eqref{defWBI} is given by
\be
\bW^{\rm def}_\a
=\bW_\a
-\zeta\widehat{\bUp}_\a\,,
\label{sol-defWBI}
\ee
where $\bW_\a$ is a regular undeformed vector multiplet field strength, see 
\eqref{undefBianchiW}--\eqref{vector-sol}, while the chiral spinor superfield 
$\widehat{\mathbf{\Upsilon}}_\a$ is such that 
\be
\label{C8}
\de^\a\widehat\bUp_\a-\deb_\ad\ov{\widehat{\bUp}}^{\ad} = 4i\,\widehat{\bL}.
\ee
The solution \eqref{sol-defWBI} is gauge invariant under 
$\d_g \widehat{\bUp}_\a=\widehat{\bW}_\a$
and $\d_g\bW_\a=\zeta\widehat{\bW}_\a$ for some  vector multiplet field strength $\widehat{\bW}_\a$
and there is a gauge in which $\bW_\a=0$. In other words, eq.~\eqref{C8} is not a constraint. It 
defines $\widehat{\bL}$ for any $\widehat\bUp_\a$.

The component fields of a $\bW^{\rm def}_\a$ 
coincide with the ones of a chiral spinor multiplet, $(\l,C,D,\cB_{ab},\chi)$, and simply arise as
\bea
&\l_\a:=\bW^{\rm def}_\a|_{\q=0} \,,
\qquad
\ov\l^\ad=\ov{\mathbf W}^{\rm def}{}^\ad|_{\q=0} \,,\qquad
D=-\frac{1}{8}\big(\de^\a \bW^{\rm def}_\a+\deb_\ad \ov{\mathbf W}^{\rm def}{}^\ad\big)|_{\q=0} \,,
\crbig
&
\cB_{ab}=\bB_{ab}|_{\q=0} \,, \qquad
\bB_{ab}:=\frac{\ri}{2}\Big(
(\s_{ab})^{\a\b}\de_{\a}\bW^{\rm def}_\b
-(\ov{\s}_{ab})_{\ad\bd}\deb^\ad\ov{\mathbf W}^{\rm def}{}^\bd
\Big) \,,
\eea
while $(C,\chi_\a,\ov\chi^\ad)$, as well as the component field strength 
$E_a= {1\over3!}\,\epsilon_{abcd}\de^b\bB^{cd}|_{\q=0}$, 
are
\be
C=\widehat{\bL}|_{\q=0}\,,\quad
\chi_\a=\de_\a\widehat{\bL}|_{\q=0} \,,\quad
\ov\chi^\ad=\deb^\ad\widehat{\bL}|_{\q=0} \,,\quad 
E_a=\frac{1}{4}(\ov\s_a)^{\ad\a}[\de_\a,\deb_\ad]\widehat{\bL}|_{\q=0} \,.
\label{compL-0}
\ee
Local $SU(2,2|1)$ transformations of a chiral spinor multiplet
can be straightforwardly derived from superspace 
and coincide (up to notation) with the ones presented in Appendix \ref{AppB}.

The gauge-fixing conditions of dilatations, ${\cal S}$-supersymmetry and special conformal transformations
are as in eqs.~\eqref{newgaugefix}, but their formulation in the curved superspace approach reads
\be
\widehat\bL = {1\over\kappa^2} \,,\qquad\qquad
{\bB}_M=0 \,,
\label{gaugeL-2}
\ee
where ${\bB}_M$ is the dilatation connection which
is pure gauge for special (super)conformal transformations \cite{Butter}.
With this gauge fixing, the residual local transformations are super-diffeomorphisms, Lorentz, 
and $U(1)_R$. The last two define the structure group of the  off-shell new-minimal Poincar\'e  
supergravity geometry which is described by the covariant derivatives
\begin{align}
\cD_A
= E_A{}^M \Big(
\pa_M
- \frac{1}{2} {\mathbf \Omega}_M{}^{a b} M_{ ab}
- {\mathbf{A}}_M \mathbb A
\Big) ,
\label{derivativesNewMinimal}
\end{align}
with ${\mathbf \Omega}_M{}^{a b}$ and ${\mathbf{A}}_M$ the Lorentz and $U(1)_R$ connections, respectively.
The geometry of $\cD_A$, originally constructed in \cite{Muller:1985vga},
can be derived by gauge fixing the $\de_A$ derivatives \cite{Butter}.
In the gauge \eqref{gaugeL-2}, and in terms of the $\cD_A$ derivatives, 
the Bianchi identity \eqref{defWBI} turns into
\be
\cD^\a\bW^{\rm def}_\a-\cDB_\ad\ov\bW^{\rm def}{}^\ad=-4i\hat{\zeta}~,
\qquad\hat{\zeta}:=\frac{\zeta}{\kappa^2} \,,
\label{defWBI-gf}
\ee
where $\hat{\zeta}$ is a constant, $\cD_A\hat{\zeta}=\pa_M\hat{\zeta}=0$.
This is the curved analog of the deformation of a Maxwell multiplet in flat superspace,
eq.~\eqref{defcSp}.

The superspace action for a deformed vector multiplet in the new-minimal supergravity background
is then given by
\bea
S_M
&=& 
- {1\over2} \Im \Bigl[\widetilde\tau\int d^4x\, d^2\theta\, \cE\, (\bW^{\rm def})^2 \Bigr] \,,\qquad
\widetilde\tau=\frac{i }{{\tilde g}^2}+\vartheta \,.
\label{magneticSuperspace}
\eea
It is dual to a vector multiplet action with a Fayet-Iliopoulos term, precisely as shown in section \ref{EMdualitySUGRA}.
It is illustrative to show how the argument given in section \ref{global-susy} 
for the global case extends to curved superspace. Instead of expression \eqref{magneticSuperspace},
one starts from the  action
\bea\label{cspaction-2}
S=-\frac{1}{2} \Imag \Big[ \widetilde\tau \int d^4x\,d^2 \theta\,\cE\, \bUp^2 \Big]
- {i\over 2}\int d^4x\,d^2\q d^2\qb\,E\, 
\mathbf U\left(\de^\alpha \bUp_\alpha - \deb_\dalpha \ov{\bUp}^\dalpha + 4i\, \zeta\widehat\bL\right) \,,
\eea
where  $\bUp_\alpha$ is a chiral spinor superfield with weights $(3/2,1)$ and
$\mathbf U$ is a zero-weight unconstrained real scalar.
Eliminating $\mathbf U$ imposes the deformed Bianchi identity \eqref{defWBI} on $\bUp_\alpha$
and, with the identification $\bUp_\alpha=\bW_\a^{\rm def}$,
one obtains the ``magnetic" action \eqref{magneticSuperspace}.
Integrating by parts and redefining a full curved superspace as a (anti-)chiral superspace 
integral,\footnote{Given a real lagrangian superfield $\mathscr{L}$ of conformal weights $(2,0)$,
 the full superspace integral is related to the (anti-)chiral 
superspace action  as  \cite{Butter}
\begin{align}
\int d^4x\, d^4\theta\, E\, \mathscr{L}
=-\frac{1}{4} \int d^4x\, d^2\theta\, \cE\, \ov\nabla^2\mathscr{L}
=-\frac{1}{4} \int d^4x\, d^2\qb\, \overline{\cE}\, \nabla^2\mathscr{L} \,.
\label{chiral-useful}
\end{align}
Local $SU(2,2|1)$ invariants can be manipulated by using the  rule for integration by parts \cite{Butter}.} 
the action \eqref{cspaction-2} proves to be equivalent to 
\be\label{cspaction-2.b}
S = -\frac{1}{2} \Imag \left[ \int d^4x\,d^2 \theta\,\cE \left(\widetilde\tau \bUp^2
- \hf\bUp^\alpha \deb^2\de_\alpha\mathbf U \right)\right]
+ 2\, \zeta \int d^4x\,d^2\q d^2\qb\,E\, \widehat{\bL}\mathbf U \,.
\ee
Eliminating the unconstrained $\bUp$ leads firstly to
\bea
\bUp_\a =-\frac{1}{\widetilde\tau}\bW_\a \,,\qquad
\bW_\a:=-\frac{1}{4} \deb^2\de_\alpha\mathbf U \,,
\eea
an undeformed abelian vector multiplet field strength, and secondly to the "electric" action
\be
S_E = -\frac{1}{2} \Imag \left[ \tau\int d^4x\,d^2 \theta\,\cE \bW^2\right]
+2\, \zeta \int d^4x\,d^2\q d^2\qb\,E\, \widehat{\bL}\mathbf U \,,\qquad
\tau=-\frac{1}{\widetilde\tau} \,,
\ee
where the second term is the curved superspace description for a standard Fayet-Iliopoulos term 
in new-minimal supergravity.



   \bibliography{ADJT_ref}
   \bibliographystyle{JHEP} 
  
\end{document}